%% file: main.tex
\documentclass[sigconf, nonacm]{acmart}
\makeatletter
\patchcmd{\@mkauthors@iii}
  {\lineskip=1pc\relax}
  {\lineskip=0.5pc\relax}
  {}{}
\makeatother

\newcommand\vldbdoi{XX.XX/XXX.XX}
\newcommand\vldbpages{XXX-XXX}
\newcommand\vldbvolume{14}
\newcommand\vldbissue{1}
\newcommand\vldbyear{2020}
\newcommand\vldbauthors{\authors}
\newcommand\vldbtitle{\shorttitle} 
\newcommand\vldbavailabilityurl{URL_TO_YOUR_ARTIFACTS}
\newcommand\vldbpagestyle{plain} 
\usepackage{enumitem}
\usepackage{algorithm}
\usepackage{algpseudocode}
\usepackage{setspace}
\usepackage{placeins}
\begin{document}
\title{BCTuner: LLM-Guided Monte Carlo Tree Search for Efficient Blockchain Knob Tuning}

%%
%% The "author" command and its associated commands are used to define the authors and their affiliations.
% \affiliation{%
%   \state{Ireland}

\author{Yaoyi Deng}
% \orcid{0000-0002-1825-0097}
\affiliation{%
  \institution{SKLCCSE Lab, Beihang University, China}
  % \streetaddress{1 Th{\o}rv{\"a}ld Circle}
  % \city{Hekla}
  % \country{Iceland}
}
\email{dengyaoyi@buaa.edu.cn}

\author{Chongyang Tao}
% \orcid{0000-0002-1825-0097}
\affiliation{%
  \institution{SKLCCSE Lab, Beihang University, China}
  % \streetaddress{1 Th{\o}rv{\"a}ld Circle}
  % \city{Hekla}
  % \country{Iceland}
}
\email{chongyang@buaa.edu.cn}

\author{Mingxuan Li}
% \orcid{0000-0002-1825-0097}
\affiliation{%
  \institution{People’s Public Security University of China}
  % \streetaddress{1 Th{\o}rv{\"a}ld Circle}
  % \city{Hekla}
  % \country{Iceland}
}
\email{20236990@ppsuc.edu.cn}

\author{Xuelian Lin}
% \orcid{0000-0002-1825-0097}
\affiliation{%
  \institution{SKLCCSE Lab, Beihang University, China}
  % \streetaddress{1 Th{\o}rv{\"a}ld Circle}
  % \city{Hekla}
  % \country{Iceland}
}
\email{linxl@buaa.edu.cn}

\author{Han Sun}
% \orcid{0000-0002-1825-0097}
% \makebox[0.4\textwidth][c]
\affiliation{%
  \institution{SIC, Tsinghua }
  % \streetaddress{1 Th{\o}rv{\"a}ld Circle}
  \city{University, China}
  % \country{Iceland}
}
\email{sunhan96@tsinghua.edu.cn}

\author{Mingchao Wan $^{*}$}
% \orcid{0000-0002-1825-0097}
\affiliation{%
  \institution{BAEC}
  % \institution{Beijing Academy of Blockchain and Edge Computing, China}
  % \streetaddress{1 Th{\o}rv{\"a}ld Circle}
  \city{Beijing, China}
  % \country{Iceland}
}
\email{chainmaker@baec.org.cn}

% \author{Mingchao Wan $^{*}$}
% % \orcid{0000-0002-1825-0097}
% \affiliation{%
%   \institution{BAEC}
%   % \institution{Beijing Academy of Blockchain and Edge Computing, China}
%   % \streetaddress{1 Th{\o}rv{\"a}ld Circle}
%   \city{Beijing, China}
%   % \country{Iceland}
% }
% \email{chainmaker@baec.org.cn}

\author{Shuai Ma $^{*}$}
% \author{-}
% \orcid{0000-0002-1825-0097}
\affiliation{%
  \institution{SKLCCSE Lab, Beihang University, China}
  % \streetaddress{1 Th{\o}rv{\"a}ld Circle}
  % \city{Hekla}
  % \country{Iceland}
}
\email{mashuai@buaa.edu.cn}

% \author{-}
% \orcid{0000-0002-1825-0097}
% \affiliation{%
%   \country{Iceland}

% \orcid{0000-0001-5109-3700}
% \affiliation{%
% \email{vb@rocquencourt.com}

% \author{J\"org von \"Arbach}
% \affiliation{%
%   \institution{University of T\"ubingen}
%   \city{T\"ubingen}
%   \country{Germany}
% }
% \email{jaerbach@uni-tuebingen.edu}
% \email{myprivate@email.com}
% \email{second@affiliation.mail}

% \author{Wang Xiu Ying}
% \author{Zhe Zuo}
% \affiliation{%
%   \institution{East China Normal University}
%   \city{Shanghai}
%   \country{China}
% }
% \email{firstname.lastname@ecnu.edu.cn}

% \author{Donald Fauntleroy Duck}
% \affiliation{%
%   \institution{Scientific Writing Academy}
%   \city{Duckburg}
%   \country{Calisota}
% }
% \affiliation{%
%   \institution{Donald's Second Affiliation}
%   \city{City}
%   \country{country}
% }
% \email{donald@swa.edu}

% \author{Wang Xiu Ying}
% \author{Zhe Zuo}
% \affiliation{%
%   \institution{East China Normal University}
%   \city{Shanghai}
%   \country{China}
% }
% \email{firstname.lastname@ecnu.edu.cn}

% \author{Wang Xiu Ying}
% \author{Zhe Zuo}
% \affiliation{%
%   \institution{East China Normal University}
%   \city{Shanghai}
%   \country{China}
% }
% \email{firstname.lastname@ecnu.edu.cn}

\begin{abstract}

Knob tuning plays a critical role in improving the performance of permissioned blockchains. 
However, efficient tuning remains challenging due to the architectural complexity of blockchains and the semantic gap between knob-specific logic and the numerical optimization requirements of tuning tools.
% However, efficient tuning remains challenging because blockchain systems expose heterogeneous knobs across multiple components, while the knowledge needed to tune them is not readily usable by blockchain tuning tools. 
In addition, configuration changes are often coupled across different stages of the transaction pipeline, making their performance impact difficult to isolate and predict. Since each trial requires deployment and distributed benchmarking, ineffective exploration incurs substantial cost. These challenges motivate BCTuner, a Large Language Model (LLM)-guided framework that combines knowledge-guided reasoning with structured search.
BCTuner organizes multi-source tuning knowledge to support LLM-based reasoning over knob semantics, constraints, and deployment context. 
It formulates tuning as a Monte Carlo Tree Search (MCTS) process over structured action trajectories, where configurations are incrementally constructed, validated, evaluated, and refined rather than generated in one step. 
BCTuner further applies adaptive pruning to discard infeasible or low-potential branches before system evaluation. 
We evaluate BCTuner on Hyperledger Fabric and ChainMaker under diverse workloads and network settings. 
Experimental results show that BCTuner achieves up to 211.38\% throughput improvement over default configurations and outperforms the state-of-the-art blockchain tuning method by up to 20\% in performance, while requiring up to $8\times$ fewer interactions with the blockchain system.

\end{abstract}

\maketitle

\pagestyle{\vldbpagestyle}
\begingroup\small\noindent\raggedright\textbf{PVLDB Reference Format:}\\
\vldbauthors. \vldbtitle. PVLDB, \vldbvolume(\vldbissue): \vldbpages, \vldbyear.\\
\href{https://doi.org/\vldbdoi}{doi:\vldbdoi}
\endgroup
\begingroup
\renewcommand\thefootnote{}\footnote{\noindent
$^{*}$Corresponding author.

This work is licensed under the Creative Commons BY-NC-ND 4.0 International License. Visit \url{https://creativecommons.org/licenses/by-nc-nd/4.0/} to view a copy of this license. For any use beyond those covered by this license, obtain permission by emailing \href{mailto:info@vldb.org}{info@vldb.org}. Copyright is held by the owner/author(s). PubliBCTunerion rights licensed to the VLDB Endowment. \\
\raggedright Proceedings of the VLDB Endowment, Vol. \vldbvolume, No. \vldbissue\ %
ISSN 2150-8097. \\
\href{https://doi.org/\vldbdoi}{doi:\vldbdoi} \\
}\addtocounter{footnote}{-1}\endgroup

\ifdefempty{\vldbavailabilityurl}{}{
\vspace{.3cm}
\begingroup\small\noindent\raggedright\textbf{PVLDB Artifact Availability:}\\
The source code, data, and/or other artifacts have been made available at \url{\vldbavailabilityurl}.
\endgroup
}

\input{1-intro}
\input{2-related}
\input{3-method}

\input{4-exp}

\FloatBarrier
\vspace{-0mm}
\section{Conclusion}
In this paper, we studied the problem of automatic configuration tuning for permissioned blockchains, which is inherently challenging due to the high-dimensional configuration space, complex dependencies among system knobs, and the high cost of evaluation. To address these challenges, we proposed BCTuner, an LLM-guided framework that integrates knowledge-driven reasoning with structured search for efficient tuning. 
The key idea of BCTuner is to formulate configuration tuning as a structured and iterative decision process, rather than a black-box trial-and-error approach or one-shot configuration generation. By incorporating multi-source knowledge and leveraging Monte Carlo Tree Search with pruning, BCTuner effectively guides the search toward high-quality configurations while significantly reducing unnecessary evaluations, resulting in more efficient optimization in complex configuration spaces. We implemented BCTuner and evaluated it on Hyperledger Fabric and ChainMaker under diverse workloads and network settings. The results show that BCTuner consistently achieves substantial performance improvements. 

\begin{acks}
 This work was supported by the [...] Research Fund of [...] (Number [...]). Additional funding was provided by [...] and [...]. We also thank [...] for contributing [...].
\end{acks}

\FloatBarrier
\clearpage
\bibliographystyle{ACM-Reference-Format}
% \bibliography{sample}
\bibliography{BCTuner}

\end{document}

%% file: 1-intro.tex
\section{Introduction}
Permissioned blockchains ~\cite{androulaki2018hyperledger, Taxonomy} have been widely adopted in enterprise settings as trusted infrastructure for decentralized transaction processing, supporting applications such as financial services ~\cite{ripple, bis_mbridge}, supply chain management ~\cite{gsbn, wetrade}, and digital asset management ~\cite{coinbase, jpm_tokenization}. Similar to distributed transactional systems, they handle large volumes of concurrent transactions while ensuring consistency and reliability~\cite{UntanglingBlockchain, ruan2021blockchains}. However, unlike traditional distributed databases, permissioned blockchains introduce additional system-level complexities, including consensus protocols for transaction ordering, cryptographic verification, and execution models that require transaction validation and replay across multiple nodes. These mechanisms are essential for trust and fault tolerance but incur computation, communication, and coordination overhead. 
As a result, the throughput of permissioned blockchains is constrained, making performance optimization a critical concern. Prior work, such as Athena~\cite{10.14778/3579075.3579076}, has shown that effective knob tuning can substantially improve blockchain performance, highlighting the need for systematic and efficient tuning.

Knob tuning is a general optimization problem in complex systems, where high-dimensional configuration spaces make finding an optimal configuration NP-hard~\cite{sullivan2004using}. Although automated tuning has been widely studied for databases and distributed systems, blockchain knob tuning poses additional challenges.
(1)~\textit{Semantic and contextual complexity.}
Blockchain systems expose heterogeneous knobs across consensus, networking, execution, validation, and resource management. Their types, units, ranges, and special values vary, and their effects depend on hardware resources, node roles, network topology, and system behavior. Thus, blockchain tuning requires semantic and contextual reasoning beyond numerical search.
(2)~\textit{Cross-stage dependencies.}
Permissioned blockchains process transactions through proposal execution, endorsement, ordering, validation, and commit. Knobs across these stages are coupled through transaction flow, consensus communication, network propagation, and resource contention, so local adjustments may shift bottlenecks across stages and make isolated tuning insufficient.
(3)~\textit{High evaluation cost.}
Each configuration trial requires system deployment or reconfiguration and distributed benchmarking, incurring substantial time and resource costs. Ineffective exploration therefore becomes prohibitive.
These challenges motivate tuning approaches that incorporate blockchain-specific knowledge, reason over cross-stage dependencies, and reduce costly evaluations.

Existing approaches to knob tuning can be broadly categorized into heuristic-based, learning-based, and, more recently, LLM-based methods. Early heuristic approaches~\cite{pgtune2014, ansel2014opentuner, zhu2017bestconfig} rely on manually designed rules or search heuristics, but they are often limited in high-dimensional configuration spaces and difficult to adapt to new systems. Learning-based methods improve tuning efficiency by modeling the relationship between configurations and performance. Bayesian optimization-based approaches~\cite{duan2009tuning, van2017automatic, zhang2021restune, cereda2021cgptuner, LlamaTune, zhang2022towards} reduce evaluations through surrogate modeling, reinforcement learning-based methods~\cite{li2019qtune, zhang2021cdbtune, wang2021udo, ge2021watuning, cai2022hunter, 10.14778/3579075.3579076} formulate tuning as sequential decision making, and deep learning-based approaches~\cite{zhang2023efficient, tan2019ibtune, bianchi2025db2une, van2021inquiry} learn workload--configuration relationships with neural models. However, these methods are primarily data-driven, relying on training data, surrogate models, or repeated system interactions to guide the search, while largely ignoring blockchain-specific knob semantics, deployment context. This makes them inefficient in blockchain settings where each candidate configuration requires costly deployment and benchmarking.

Recent advances in LLMs have demonstrated strong semantic understanding and structured reasoning capabilities~\cite{yang2025qwen3, achiam2023gpt, guo2025deepseek}, making them promising for configuration tuning tasks that rely on textual knob descriptions and domain rules. Existing studies explore this direction by using LLMs or textual knowledge to assist search space construction and knob selection~\cite{trummer2022db, lao2025gptuner, yan2025mctuner}, while the optimization process still largely depends on iterative trial-and-error evaluations. Other methods directly use LLMs to recommend configurations~\cite{huang2024e2etune, giannakouris2025lambda}, but their reliance on offline training data and database-specific representations limits their adaptability to new systems and deployment environments. More recently, LLM-agent-based tuning has moved beyond static recommendation by decomposing tasks, incorporating runtime feedback, and refining decisions iteratively. AgentTune~\cite{li2025agenttune} represents an early effort in this direction, using workload analysis and knob selection to guide iterative database tuning. However, such methods may struggle with tightly coupled systems such as blockchains, where a single knob adjustment can affect multiple components and transaction stages. In these environments, effective tuning requires reasoning over intermediate decisions and their long-term effects, rather than merely generating complete candidate configurations.
Therefore, existing LLM-based tuning methods remain insufficient for permissioned blockchains, where tuning requires reasoning over heterogeneous knobs, cross-stage transaction-pipeline dependencies, and costly system evaluations. This motivates a tuning framework that can organize blockchain-specific knowledge and reason over intermediate tuning decisions during configuration construction, evaluation, and refinement.

% Existing LLM-based tuning studies explore this direction in different ways. Some methods use LLMs or textual knowledge to assist search space construction or knob selection~\cite{trummer2022db, lao2025gptuner, yan2025mctuner}, but the optimization process still relies on iterative trial-and-error evaluations. Other methods directly predict configurations from workload features or generate full configurations with LLMs~\cite{huang2024e2etune, giannakouris2025lambda}, reducing online search but often relying on large-scale offline data, workload-specific assumptions, or one-shot generation. Recent agent-based methods further introduce iterative refinement~\cite{li2025agenttune}, yet they are designed around database-specific abstractions such as SQL workloads and execution-level metrics. 

% Therefore, existing LLM-based tuning methods cannot be directly applied to permissioned blockchains, where tuning requires reasoning over heterogeneous blockchain knobs, cross-stage transaction-pipeline dependencies, and costly system evaluations. This motivates a tuning framework that can systematically organize blockchain-specific knowledge, refine configurations iteratively, and guide search toward promising configurations with fewer evaluations.

To address these challenges, we propose BCTuner, an LLM-guided tuning framework that integrates knowledge-guided reasoning with structured search. BCTuner organizes multi-source tuning knowledge to ground LLM-based reasoning in knob semantics, constraints, and deployment context. Instead of treating the LLM as a one-shot configuration generator, BCTuner embeds LLM-guided actions into an MCTS process, where configurations are incrementally constructed, validated, evaluated, and refined along structured action trajectories. During this process, adaptive pruning uses validation results and runtime feedback to discard infeasible or low-potential branches before costly system evaluation. By jointly combining knowledge organization, action-based refinement, and search pruning, BCTuner shift blockchain knob tuning from black-box configuration prediction into knowledge-guided configuration refinement along action trajectories.

We summarize our contributions as follows:
% \vspace{-5mm}

\begin{itemize}
\item \textbf{LLM-based tuning agent for blockchain systems.} We propose BCTuner, to the best of our knowledge the first LLM-based framework for blockchain knob tuning, enabling structured and iterative configuration optimization.
\item \textbf{Knowledge-guided tuning with LLMs.}
We leverage LLMs to extract and use knob and system knowledge, shifting blockchain knob tuning from data-driven exploration to knowledge-guided configuration search. 
\item \textbf{Action-based tuning formulation.}
We decompose block\-chain knob tuning into a set of tuning actions that progressively refine configurations across different stages of the tuning trajectory.
\item \textbf{Efficient search with MCTS and pruning.}
We employ MCTS with tailored pruning strategies to guide exploration of the configuration space, reducing unnecessary trials and the number of expensive evaluations.
\item \textbf{Extensive evaluation on real blockchain platforms.}
BCTuner achieves up to \textbf{211.38\%} throughput improvement and is up to \textbf{8$\times$} faster than state-of-the-art methods.
\end{itemize}

% The remainder of this paper is organized as follows. Section~2 reviews related work. Section~3 presents the BCTuner framework and problem formulation. Section~4 describes the LLM-guided MCTS-based configuration generation method. Section~5 presents the experimental evaluation. Finally, Section~6 concludes the paper.

%% file: 2-related.tex
\section{BACKGROUND AND RELATED WORK}

\subsection{Blockchain Knob Tuning}

Permissioned blockchain systems, such as Hyperledger Fabric~\cite{androulaki2018hyperledger} and ChainMaker~\cite{chainmaker2024}, are widely used in enterprise applications where transaction throughput and reliability are critical. These systems expose configuration knobs across multiple components, including transaction submission, endorsement, ordering, validation, commit, storage, networking, and resource management. These knobs control system behaviors such as batching, concurrency, communication, state access, and resource allocation. Blockchain knob tuning aims to select appropriate knob values under a given workload, network topology, and hardware environment to improve end-to-end transaction performance. 

Unlike traditional database systems, permissioned blockchains process transactions through a multi-stage pipeline. A transaction typically goes through proposal execution, endorsement, ordering, validation, and commit before it is finalized. Each stage is implemented by different components and controlled by different groups of knobs, but their effects are not independent. For example, increasing the block size or batch timeout at the ordering layer may improve batching efficiency by packing more transactions into each block. However, larger blocks also increase the amount of validation, state access, and commit work performed by peers, and may shift the bottleneck from ordering to validation or storage. Therefore, blockchain performance is a pipeline-level outcome determined by interactions among consensus, execution, networking, storage, and resource-management components.

Taken together, these properties make blockchain knob tuning different from conventional DBMS tuning. Effective tuning must account for heterogeneous knob semantics, deployment-specific context, cross-stage effects along the transaction pipeline, and the high cost of evaluating configurations on a deployed blockchain network. This makes purely black-box or trial-and-error tuning less suitable, motivating the need for knowledge-aware and evaluation-efficient tuning methods.

\subsection{Traditional Database Knob Tuning}
Traditional database knob tuning methods can be broadly divided into four categories. 
(1) \textit{Heuristic-based methods}, such as PGTune~\cite{pgtune2014}, OpenTuner~\cite{ansel2014opentuner}, and BestConfig~\cite{zhu2017bestconfig}, rely on expert-designed rules, enabling quick guidance but requiring substantial domain expertise and offering limited generalization across workloads and environments.
(2) \textit{Bayesian optimization-based methods}, including iTuned~\cite{duan2009tuning}, OtterTune~\cite{van2017automatic}, ResTune~\cite{zhang2021restune}, CGPTuner~\cite{cereda2021cgptuner}, LlamaTune~\cite{LlamaTune}, and OnlineTune~\cite{zhang2022towards}, use surrogate models to approximate the performance landscape and guide search. However, they typically require high-quality historical data, and their effectiveness can degrade under workload shifts, limiting generalization to unseen scenarios.
(3) \textit{Reinforcement learning-based methods}, such as Qtune~\cite{li2019qtune}, CDBTune~\cite{zhang2021cdbtune}, UDO~\cite{wang2021udo}, WATuning~\cite{ge2021watuning}, Hunter~\cite{cai2022hunter}, and Athena~\cite{10.14778/3579075.3579076}, formulate tuning as sequential decision-making and learn policies for configuration adjustment. Although they can adapt to dynamic environments, they often require unstable training and extensive system interactions, leading to high tuning overhead.
(4) \textit{Deep learning-based methods}, including iBTune~\cite{tan2019ibtune}, OpAdviser~\cite{zhang2023efficient}, and Db2une~\cite{bianchi2025db2une}, use neural networks to model performance or guide search. These methods depend on sufficient training data and still require iterative refinement, making them less effective in data-scarce or changing environments.
Despite their differences, these approaches either rely on high-quality historical data or require extensive online interactions with the system, making efficient and robust tuning difficult in complex and dynamic environments.

\subsection{LLM-based Database Knob Tuning}
LLMs have recently shown strong capabilities in semantic understanding and reasoning, and have been applied to various database tasks such as performance diagnosis~\cite{D-Bot, 10.1145/3709663}, text-to-SQL~\cite{wang2025mac, bai2025judgesql}, and query optimization~\cite{GenRewrite, li2024llm}. These capabilities make LLMs a promising tool for configuration tuning, where understanding knob semantics and parameter interactions is essential.

Several recent studies have explored LLMs for database knob tuning. DB-BERT~\cite{trummer2022db} extracts tuning hints from textual documents, but still relies on reinforcement learning-based trial-and-error search and is sensitive to external knowledge quality. GPTuner~\cite{lao2025gptuner} uses LLMs to guide Bayesian optimization, but treats them as auxiliary components and still requires extensive iterative interactions, with limited generalization under workload shifts. $\lambda$-Tune~\cite{giannakouris2025lambda} generates full configurations via LLMs and reduces evaluation overhead through workload-specific scheduling, but is coupled with query execution characteristics and suffers from unstable single-shot generation in high-dimensional knob spaces. E2ETune~\cite{huang2024e2etune} learns workload--configuration mappings from large-scale offline data, but costly data generation and retraining limit its adaptability to new systems and dynamic environments. MCTuner~\cite{yan2025mctuner} uses LLMs for knob selection and combines space decomposition with Bayesian optimization, but still treats LLMs as preprocessing tools and requires many iterative evaluations. AgentTune~\cite{li2025agenttune} supports structured decision-making through task decomposition and iterative refinement, but relies on database-specific abstractions such as SQL workload semantics and execution-level metrics. Overall, these methods are designed for database systems and depend on workload-specific knowledge, historical data, or repeated system interactions, making them difficult to apply to permissioned blockchains with distinct execution models and limited tuning knowledge.

%% file: 3-method.tex
\section{BCTuner Framework}

\begin{figure*}[t]
\centering
% \vspace{-3mm}
\includegraphics[width=0.88\textwidth]{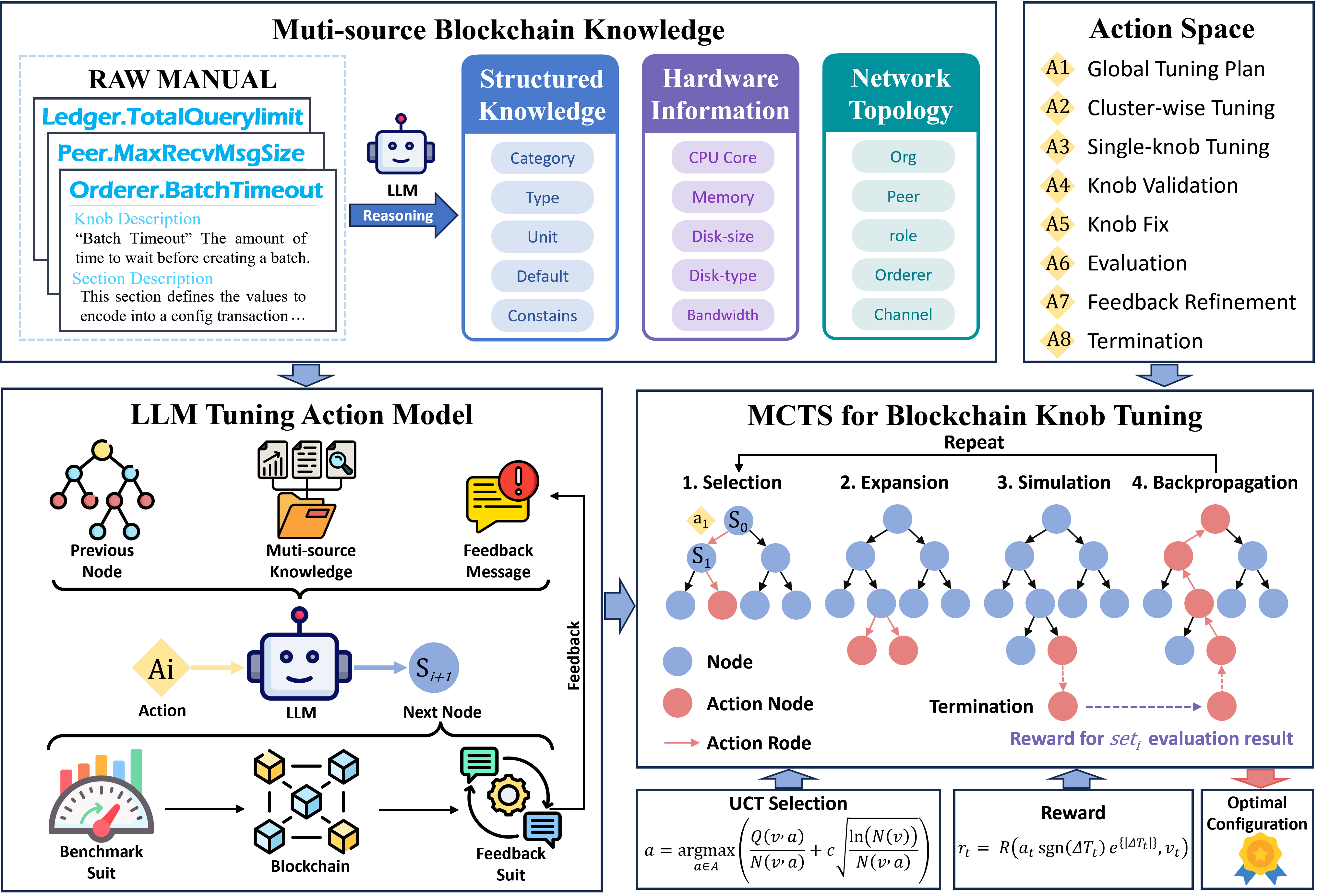}
% \vspace{-3mm}
\caption{ System Overview of BCTuner.}
% \vspace{-3mm}
\label{fig:bctuner_overview}
% \vspace{-4mm}
\end{figure*}

\subsection{Problem Formulation}
\noindent\textbf{Blockchain knob tuning.} We formally define the blockchain knob tuning task as follows. Let $\mathcal{B} = (H, N)$ denote a blockchain system, where $H$ represents the hardware environment and $N$ denotes the network topology. Let $\mathcal{K} = \{k_1, k_2, \dots, k_m\}$ denote the set of tunable knobs. Each knob $k_i$ is associated with a value range $\mathcal{V}_i$. The configuration space is defined as the Cartesian product of the value ranges of all knobs, i.e., $\mathcal{C} = \mathcal{V}_1 \times \mathcal{V}_2 \times \cdots \times \mathcal{V}_m$. A configuration is an assignment of values to all knobs, denoted as $C = (kv_1, kv_2, \dots, kv_m)$, where $kv_i$ is the value of knob $k_i$.

Given a blockchain system $\mathcal{B}$, a configuration $C$, and a smart contract workload $S$, the system performance is evaluated by a function $\mathrm{Eval}$, measured in throughput (transactions per second, TPS), denoted by $T$:
\begin{equation}
T = \mathrm{Eval}(\mathcal{B}, C, S).
\end{equation}
The goal of blockchain knob tuning is to identify an optimal configuration $C^*$ that maximizes system performance under a given hardware environment, network topology, and workload. Formally, the optimization problem is defined as:
\begin{equation}
C^* = \arg\max_{C \in \mathcal{C}} \; \mathrm{Eval}(\mathcal{B}, C, S).
\end{equation}

\noindent\textbf{Blockchain knob tuning as a Search Problem.}
% Traditional tuning approaches typically rely on machine learning techniques that require a large amount of configuration--performance data for training~\cite{lao2025gptuner, 10.14778/3579075.3579076}. However, collecting such data is expensive and time-consuming, especially in blockchain systems where each evaluation incurs substantial overhead.
Traditional tuning approaches are largely data-driven, relying on configuration--performance observations collected through iterative interactions with the target system~\cite{lao2025gptuner, 10.14778/3579075.3579076}. 
However, obtaining such observations is expensive and time-consuming in blockchain environments, where each evaluation requires deploying or reconfiguring the system and running distributed workload benchmarks.

LLMs exhibit strong capabilities in semantic understanding and reasoning. They can capture the meanings of knobs and their interdependencies, and generate informed adjustments. This motivates the use of LLMs to simulate the decision-making process of human blockchain administrators for configuration tuning.
However, directly relying on LLMs is insufficient due to the large search space and the complex coupling among knobs, which make purely heuristic or one-shot decisions unreliable. Therefore, it is necessary to incorporate a structured search mechanism to systematically explore and refine configurations.

To address the blockchain knob tuning problem, we formulate it as an LLM-guided MCTS problem. Given the large search space and expensive evaluations, MCTS provides an effective framework that balances exploration and exploitation, enabling efficient search toward high-quality configurations.

Formally, let $s$ denote a search state, where each state corresponds to a configuration $C$, and let $a \in \mathcal{A}$ denote a tuning action. We construct a search tree $\mathcal{T}$, in which each node represents a state $s$, and each edge corresponds to an action that transforms one state into another. Applying action $a_t$ to state $s_t$ yields the next state $s_{t+1} = \delta(s_t, a_t)$, where $\delta$ denotes the state transition function.

The tuning process is modeled as a sequential search over action sequences $\pi = (a_0, \dots, a_{|\pi|-1})$, where each sequence defines a path in the search tree starting from an initial state $s_0$, and $\Pi$ denotes the set of all feasible action sequences. Let $s_t^\pi$ denote the state at step $t$ along trajectory $\pi$, and let $C_t^\pi$ denote the corresponding configuration. The objective is to identify the configuration that achieves the best performance over all explored trajectories:
\begin{equation}
C^* = \arg\max_{\pi \in \Pi} \; \max_{0 \le t \le |\pi|} \; \mathrm{Eval}(\mathcal{B}, C_t^\pi, S).
\end{equation}

where for each trajectory $\pi$, the inner maximization selects the best-performing configuration along the trajectory, and the outer maximization identifies the trajectory that yields the highest performance overall. As a result, $C^*$ corresponds to the configuration that achieves the maximum throughput among all configurations explored during the search. 

\vspace{3mm}
\subsection{BCTuner overview}
\noindent\textbf{Overall Design.} Blockchain knob tuning presents an NP-hard optimization problem due to the high-dimensional configuration space and the intricate interactions among consensus, network, and execution components. These characteristics introduce several key challenges: (a) The configuration space is extremely large, and performance depends on complex and often non-linear interactions among knobs. Without sufficient understanding of knob semantics and their relationships, exploration can easily fall into ineffective or invalid regions. (b) Directly generating a full configuration in one shot is unreliable in high-dimensional knob spaces. Without structured guidance, important dependencies among knobs can be easily overlooked, leading to low-quality or even invalid configuration settings. (c) Configuration evaluation is expensive, as each candidate must be deployed and benchmarked on the system. This necessitates minimizing unnecessary trials and maximizing the utility of each evaluation.

% BCTuner addresses these challenges by introducing an LLM-guided MCTS framework for blockchain knob tuning. To address the challenge (a), BCTuner incorporates multi-source knowledge, including knob semantics and system-level characteristics, to guide the search toward valid and effective regions of the configuration space. To address the challenge (b), BCTuner introduces a structured action-based tuning process, where the LLM incrementally refines configurations through a sequence of decisions, enabling it to better capture dependencies among knobs and avoid unreliable one-shot generation. To address the challenge (c), BCTuner leverages MCTS to organize the exploration process, progressively focusing on promising configurations and reducing the number of costly benchmark evaluations. 
% \vspace{1mm}

The above challenges lead to three design requirements for effective blockchain tuning. 
BCTuner addresses the above requirements through an LLM-guided MCTS framework for blockchain knob tuning. 
First, BCTuner incorporates multi-source knowledge, including knob semantics and system-level context, to ground tuning decisions in explicit blockchain knowledge and guide the search toward valid and effective configurations. 
Second, instead of generating full configurations in a single step, BCTuner organizes tuning as a structured action-based process, where the LLM progressively refines configurations through iterative decisions. 
Third, BCTuner leverages MCTS with adaptive pruning to organize exploration and reduce unnecessary deployment and benchmarking during tuning.

The integration of knowledge-guided reasoning and structured search enables BCTuner to perform effective tuning in blockchain systems. In our experiments, BCTuner achieves up to 211.38\% performance improvement and reduces tuning time by up to $14\times$ compared with traditional approaches, demonstrating the effectiveness of this design. The approach also provides a general paradigm for applying LLMs to high-dimensional system optimization.

\vspace{2mm}
\noindent\textbf{Workflow.} Fig.~\ref{fig:bctuner_overview} illustrates the overall workflow of BCTuner, which consists of two phases: an offline knowledge preparation phase and an online tuning phase.

In the offline phase, BCTuner collects multi-source information from system manuals and runtime metadata, including knob semantics, hardware characteristics, and network topology. This information is further processed and organized into structured representations, capturing key properties such as knob roles, implicit constraints, and relationships among knobs. By transforming heterogeneous and unstructured inputs into a unified and structured form, this phase provides actionable guidance that can be consistently utilized throughout the tuning process, effectively bridging raw system knowledge and automated decision-making.

In the online phase, BCTuner formulates knob tuning as a search over the configuration space. BCTuner incrementally constructs a configuration through a sequence of reasoning actions. At each step, partial decisions are extended and refined until a complete configuration is obtained. The generated configuration is then deployed and evaluated on the blockchain system to obtain performance metrics. The evaluation results are used as feedback to guide subsequent tuning, forming a closed-loop optimization process. Through iterative generation, evaluation, and feedback, configurations are progressively improved over multiple steps, enabling effective exploration of the configuration space.

\section{LLM-Guided MCTS for Blockchain Knob Tuning}
\subsection{Multi-Source Blockchain Knowledge}
% In practice, human blockchain administrators first understand how each knob functions within the system before tuning. This understanding enables reasoning about feasible configurations and helps avoid invalid or undesirable settings. Without such knowledge, tuning becomes unguided and often leads to invalid configurations or inefficient exploration. These observations suggest that effective knob tuning requires not only exploring the configuration space, but also capturing knob semantics, constraints, and behavior.

% Based on this insight, BCTuner incorporates multi-source knowledge into the tuning process to support knowledge-aware exploration. Specifically, BCTuner leverages three types of knowledge: (1) knob-level knowledge, (2) system-level knowledge, and (3) network-level knowledge. Knob-level knowledge captures the semantics of individual knobs, including their functional effects and constraints. System-level knowledge describes hardware resources, such as CPU, memory, and storage, which affect configuration feasibility. Network-level knowledge encodes structural properties of the blockchain system, including node roles and topology, which influence knob interactions and system behavior.

In practice, human blockchain administrators first understand how each knob functions before tuning, which helps them reason about feasible configurations and avoid invalid or undesirable settings. Without such knowledge, tuning becomes unguided and may lead to invalid configurations or inefficient exploration. This suggests that effective knob tuning requires not only searching the configuration space, but also capturing knob semantics, constraints, and behavior. Based on this insight, BCTuner incorporates multi-source knowledge to support knowledge-aware exploration. Specifically, it leverages three types of knowledge: \textit{knob-level}, \textit{hardware-level}, and \textit{network-level} knowledge. Knob-level knowledge captures the semantics, functional effects, and constraints of individual knobs. Hardware-level knowledge describes resources such as CPU, memory, and storage, which affect configuration feasibility. Network-level knowledge encodes node roles and topology, which influence knob interactions and system behavior.

\vspace{1mm}
\noindent\textbf{Knob Knowledge Construction.} 
% Unlike database systems, where tuning knowledge can be collected from forums and historical workloads, 
Unlike database systems, where a mature ecosystem provides diverse sources of tuning knowledge, blockchain knob knowledge is mainly derived from official manual. This limited but authoritative source requires a structured pipeline to extract and organize tuning-relevant information.

BCTuner constructs knob knowledge in three stages. First, the LLM identifies performance-related knobs from system manuals and extracts their descriptions, covering knobs that directly or indirectly affect performance. Second, it infers key attributes of each knob, including type (e.g., integer, float, boolean, or string), unit, and special values, which provide constraints for valid configuration generation. Third, it groups knobs into clusters according to their functional roles in the blockchain execution pipeline, such as transaction processing, networking, and resource management. This LLM-based grouping captures semantic relationships between knobs and system components.

\vspace{1mm}
\noindent\textbf{Hardware and Network Knowledge Construction.} In addition to knob-level knowledge, BCTuner incorporates hardware-level and network-level Knowledge to capture the deployment environment. This Knowledge is provided by administrators and encoded in structured formats (e.g., JSON), including hardware specifications and network topology.
hardware-level knowledge captures hardware characteristics such as CPU, memory, and storage, which constrain feasible configurations. Network-level knowledge describes structural properties of the blockchain system, including node roles and connectivity, which influence knob interactions and system performance. Incorporating this Knowledge allows BCTuner to contextualize tuning decisions and avoid configurations that are incompatible with the system environment.

\vspace{1mm}
% Taken together, knob-level, hardware-level, and network-level knowledge provide a representation of the blockchain system, enabling BCTuner to reason about both the feasibility and effectiveness of configurations. By grounding the tuning process in explicit knowledge rather than treating it as a black-box optimization problem, BCTuner guides exploration toward valid and high-quality regions of the configuration space, improving efficiency and reliability.
Taken together, knob-level, hardware-level, and network-level knowledge provide a structured representation of the blockchain system. This representation allows BCTuner to assess both configuration feasibility and potential performance effects. By grounding tuning decisions in explicit system knowledge rather than treating knob tuning as black-box optimization, BCTuner guides the search toward valid and promising regions of the configuration space, improving tuning efficiency and robustness.

\subsection{LLM as the Tuning Action Model}
% BCTuner models the LLM as an action generator within an iterative tuning loop, rather than treating configuration generation as a one-shot process. At each step, the LLM produces a valid and executable action that incrementally modifies the current configuration. This design alleviates the difficulty of reasoning over high-dimensional configuration spaces in a single pass, where attention may be diluted across numerous interdependent knobs, leading to suboptimal decisions~\cite{liu2024lost,bai2024longbench}. By decomposing the tuning process into a sequence of interdependent actions, configurations are progressively refined, resulting in more stable and reliable optimization in complex search spaces. In this way, the formulation enables the analysis of knob interactions and supports structured decision-making.

% To support diverse tuning decisions, BCTuner adopts a dynamic prompt construction mechanism tailored to different action types. Instead of using a fixed prompt, it employs action-specific prompts with distinct objectives and contextual information. For each action, the prompt is constructed by incorporating the current configuration, system context, and task instructions. This design allows the LLM to focus on a subset of knobs and generate more precise and context-aware adjustments.

BCTuner models the LLM as an action generator within an iterative tuning loop, rather than treating configuration generation as a one-shot process. At each step, the LLM produces a action that incrementally modifies the current configuration. 
This design alleviates the difficulty of adjusting all knobs in a single pass, where the LLM's attention may be dispersed across numerous knobs~\cite{liu2024lost,bai2024longbench}, leading to suboptimal decisions. 
By decomposing tuning into a sequence of interdependent actions, BCTuner progressively refines configurations, supports reasoning over knob interactions, and enables structured decision-making. 

To support diverse tuning decisions, BCTuner adopts action-specific prompts instead of a fixed prompt. Each prompt is constructed from the current configuration, system context, and task instructions, with objectives and contextual information tailored to the corresponding action type. This design constrains the reasoning scope of the LLM and supports precise, context-aware adjustments.

% Formally, at step $t$, given the current state $s_t$ (corresponding to configuration $C_t$), the LLM generates an action $a_t \in \mathcal{A}$, and the next state is obtained as:
% \begin{equation}
% s_{t+1} = \mathrm{LLM}(s_t, \mathrm{Prompt}(a_t)),
% \end{equation}

% Here, $\mathrm{Prompt}(\cdot)$ denotes the instruction associated with a specific action. Under this formulation, the LLM performs reasoning-driven state transitions, analogous to an expert making tuning decisions based on the current configuration and task context.

Formally, at step $t$, given the current state $s_t$ (corresponding to configuration $C_t$), the LLM generates an action $\mathcal{A}_t \in \mathcal{A}$ conditioned on the current state and a dynamically constructed prompt, $\mathrm{type}(a_t)$ denotes the action type, and $\mathrm{Prompt}(s_t, \mathrm{type}(a_t))$ is constructed by combining the current configuration, system context, and task instructions associated with that action type. The environment then applies the action to obtain the next state:
\begin{equation}
s_{t+1} = \delta(s_t, \mathcal{A}_t = \mathrm{LLM}(\mathrm{Prompt}(s_t, \mathrm{type}(\mathcal{A}_t))),
\end{equation}
where $\delta(\cdot)$ denotes the state transition function induced by executing the action in the current state.

% This design allows different actions to operate under distinct reasoning contexts. For example, actions such as direction selection, cluster-level adjustment, knob refinement, validation, and feedback incorporation correspond to different aspects of the tuning process. By aligning prompts with the semantics of each action, BCTuner guides the LLM to perform structured reasoning within well-defined decision scopes, reducing ambiguity and improving action quality.

% Overall, the LLM performs reasoning-driven state transitions conditioned on the current state and action-specific prompts. This formulation enables the tuning process to follow structured decision-making rather than unconstrained generation, and integrates naturally with search-based optimization, where each action corresponds to a state transition that progressively steers exploration toward high-quality configurations.

This design allows different actions to operate under distinct reasoning contexts, where prompts are aligned with action semantics to guide the LLM within well-defined decision scopes. As a result, the LLM induces state transitions conditioned on the current state and action-specific prompts, enabling structured decision-making that integrates with search-based optimization.

\subsection{Tuning Action Design}
\label{sec:action_design}
% To enable structured tuning, BCTuner defines a set of actions that decompose the knob tuning process into a sequence of explicit decision steps. Instead of generating complete configurations in a single pass, tuning is formulated as a sequence of fine-grained decisions, where each action focuses on a specific aspect of optimization. At each step, an action is executed to either update the current configuration or provide guidance for subsequent decisions, progressively steering the search toward higher-quality configurations. This design reduces the difficulty of reasoning over high-dimensional and interdependent knob spaces, allowing the tuning process to evolve in a more stable and controllable manner.

% Each action corresponds to a distinct type of decision in the tuning process, including direction selection, cluster-level adjustment, knob refinement, validation, and feedback incorporation. These actions are not limited to direct knob updates, but also include operations that analyze system behavior, verify feasibility, and use performance feedback to guide subsequent decisions. By modeling these decisions as actions, BCTuner transforms configuration search into a sequence of explicit and manageable operations. This formulation improves transparency and supports more effective exploration through targeted adjustments and iterative refinement. We next describe each action type.

To enable structured tuning, BCTuner defines a set of actions that decompose configuration optimization into a sequence of decision steps. Instead of generating complete configurations in a single pass, the tuning process explores the configuration space by selecting and applying actions at each step, where actions correspond to levels of decision granularity, including Global Tuning Plan, Cluster-wise tuning, and Single-knob tuning. The configurations are complemented by Validation and Feedback-driven refinement, with MCTS guiding the selection of the next action based on the current search state. We next describe each action type.

\vspace{1mm}
% \noindent\textbf{A1. Global Tuning Plan.}
% The \textit{Direction} action determines the high-level optimization focus at each stage of tuning. 
% Given the current state, including hardware and network Knowledge $\mathcal{K}^{sys}$, 
% cluster organization $\mathcal{G}$, and untuned clusters $\mathcal{G}_t^{untuned}$, 
% this action identifies a promising optimization direction $d_t$ that is likely to improve performance. 
% Instead of modifying knob values directly, it provides coarse-grained guidance on which parts of the system to tune next.
% \begin{equation}
% d_t = \text{A1}_{plan}(\mathcal{K}^{net}, \mathcal{K}^{sys}, \mathcal{G}, \mathcal{G}_t^{untuned})
% \end{equation}
% To ensure a structured and effective tuning process, the Direction action follows a predefined priority scheme aligned with the blockchain execution pipeline. It prioritizes clusters along the transaction processing flow (e.g., gateway, execution, ordering, and validation/commit), followed by system-level components such as resource management, network communication, and monitoring. It also incorporates feedback from previous evaluations and considers dependencies among clusters to identify bottlenecks and focus on impactful optimization opportunities. 
% The resulting direction $d_t$ guides subsequent actions, narrowing the search space and enabling more targeted exploration.

\noindent\textbf{A1. Global Tuning Plan.}
The \textit{Global Tuning Plan} action determines the high-level optimization plan at each stage of tuning.
Given the current state, including hardware and network knowledge $\mathcal{K}^{sys}$, cluster organization $\mathcal{G}$, and untuned clusters $\mathcal{G}_t^{untuned}$, this action generates a planning decision $p_t$ that specifies which parts of the system should be prioritized in subsequent tuning steps. This action operates at a global level, determining how optimization efforts are allocated across system components. Formally, the planning decision is defined as:
\begin{equation}
p_t = \mathcal{A}_{plan}(\mathcal{K}^{net}, \mathcal{K}^{sys}, \mathcal{G}, \mathcal{G}_t^{untuned}).
\end{equation}
To ensure a structured tuning process, the planning decision considers the blockchain execution pipeline, prioritizing clusters along the transaction processing flow (e.g., gateway, execution, ordering, and validation/commit), as well as system-level components such as resource management and network communication.
It also incorporates feedback from previous evaluations and captures dependencies among clusters to identify potential bottlenecks. The resulting plan $p_t$ guides subsequent actions by narrowing the search space and determining the focus of future tuning steps.
\vspace{1mm}

\noindent\textbf{A2. Cluster-wise Tuning.}
The \textit{Cluster-wise Tuning} action performs coarse-grained configuration updates by selecting a cluster and jointly tuning its associated knobs.
Given the planning decision $p_t$, system and network knowledge $\mathcal{K}^{sys}$, knob-level knowledge $\mathcal{K}^{knob}$, cluster definitions $\mathcal{G}$, the set of untuned clusters $\mathcal{G}_t^{un}$, and the set of previously tuned knobs $\mathcal{C}_{t-1}^{tuned}$, this action selects a cluster $g_t \in \mathcal{G}_t^{un}$ and generates coordinated updates for the knobs within the selected cluster.
\begin{equation}
C_t^{(g_t)} = \mathcal{A}_{cluster}(p_t, \mathcal{K}^{sys}, \mathcal{K}^{knob}, \mathcal{G}, \mathcal{G}_t^{un}, \mathcal{C}_{t-1}^{tuned})
\end{equation}
where $C_t^{(g_t)}$ denotes the sub-configuration corresponding to cluster $g_t$, representing the assigned values of all knobs within the selected cluster at step $t$.
Once a cluster is selected, the action retrieves its associated knobs based on the predefined cluster--knob mapping and generates updated values while maintaining consistency with previously tuned knobs. 
By jointly tuning multiple related knobs, this action captures their semantic relationships and functional dependencies, producing coordinated configurations rather than independent updates.
Operating at the cluster level, it provides an effective intermediate step between global planning and fine-grained knob-level tuning. 
The resulting sub-configuration $C_t^{(g_t)}$ serves as the basis for subsequent actions such as refinement or validation, enabling targeted and progressive optimization.
\vspace{1mm} 

% \noindent\textbf{A3. Single-knob Tuning.}
% The \textit{Knob Fine-Tuning} action performs fine-grained adjustment by selecting a critical knob and refining its value based on the current system state. 
% Given the optimization direction $p_t$, system and network Knowledge $\mathcal{K}^{sys}$, knob-level knowledge $\mathcal{K}^{knob}$, 
% the previously tuned configuration $\mathcal{C}_{t-1}^{tuned}$, 
% this action identifies a critical knob $k_t$ that requires further refinement.
% \begin{equation}
% (k_t, v_t) = \text{A3}_{single}(p_t, \mathcal{K}^{sys}, \mathcal{K}^{knob}, \mathcal{C}_{t-1}^{tuned})
% \end{equation}
% where $k_t$ denotes the selected knob and $v_t$ denotes its refined value at step $t$. Unlike cluster-level tuning, which focuses on coordinated adjustments over multiple knobs, this action targets precise optimization at the individual knob level. 
% When selecting the knob, the action considers both tuned and untuned knobs, as well as their interactions with previously adjusted knobs. 
% Rather than enforcing strict consistency with the existing configuration, it focuses on identifying critical knobs that are not fully captured during cluster-level tuning and require further refinement. 
% By targeting such under-optimized knobs, this action complements cluster-level adjustments and enables more precise performance improvement. 
% \vspace{1mm}

\noindent\textbf{A3. Single-knob Tuning.}
The \textit{Single-knob Tuning} action performs fine-grained configuration updates by selecting a critical knob and refining its value based on the current state.
Given the plan $p_t$, system and network knowledge $\mathcal{K}^{sys}$, knob-level knowledge $\mathcal{K}^{knob}$, and the previously tuned configuration $\mathcal{C}_{t-1}^{tuned}$, this action selects a knob $k_t$ and generates an updated value $v_t$ for that knob.
\begin{equation}
(k_t, v_t) = \mathcal{A}_{single}(p_t, \mathcal{K}^{sys}, \mathcal{K}^{knob}, \mathcal{C}_{t-1}^{tuned})
\end{equation}
where $k_t$ denotes the selected knob and $v_t$ denotes its updated value at step $t$.
Unlike cluster-wise tuning, which jointly adjusts multiple related knobs, this action focuses on refining a single knob to capture fine-grained improvements.
The action identifies under-optimized knobs that require further refinement and updates them while considering their interactions with previously tuned parameters.
By operating at the individual knob level, this action complements cluster-wise tuning and enables more precise performance improvement.
\vspace{1mm}

\noindent\textbf{A4. Knob Validation.}
The \textit{Knob Validation} action evaluates the rationality and feasibility of the current tuned configuration. 
Given the current configuration $\mathcal{C}_{t-1}^{tuned}$, system and network Knowledge $\mathcal{K}^{sys}$ and knob-level knowledge $\mathcal{K}^{knob}$, this action assesses whether the configuration is valid under the given system context.
\begin{equation}
(v_t, \mathcal{E}_t^{val}) = \mathcal{A}_{validation}(\mathcal{C}_{t-1}^{tuned}, \mathcal{K}^{sys}, \mathcal{K}^{knob})
\end{equation}
where $v_t \in \{0,1\}$ indicates whether the configuration is valid, and $\mathcal{E}_t^{val}$ denotes the set of detected issues, including constraint violations, format errors, and logical inconsistencies. This action focuses on identifying configurations that violate constraints or exhibit internal inconsistencies. 
It checks whether knob values remain within valid ranges, follow correct formats, and are compatible with related configurations under the current system context, 
filtering out invalid or conflicting configurations. 
The output $\mathcal{E}_t^{err}$ explicitly identifies problematic knobs, enabling targeted correction in subsequent steps.
\vspace{1mm}

\begin{table}[t!]
\centering
\caption{Action Ordering in BCTuner}
% \vspace{-3mm}
\label{tab:action_transition}
\renewcommand{\arraystretch}{1.15} 
\begin{tabular*}{\columnwidth}{@{\hspace{18pt}}c@{\extracolsep{\fill}}c@{\hspace{18pt}}}
\toprule
\textbf{Current Action} & \textbf{Valid Next Actions} \\
\midrule
Root & A1 \\
A1 Global Tuning Plan & A2, A3 \\
A2 Cluster-wise Tuning & A2, A3, A4, A6 \\
A3 Single-knob Tuning & A2, A3, A4 \\
A4 Knob Validation & A2, A3, A5, A6, A8 \\
A5 Knob Fix & A4 \\
A6 Performance Evaluation & A1, A7, A8 \\
A7 Feedback-driven Refinement & A8 \\
A8 Terminal & -- \\
\bottomrule
\end{tabular*}
% \vspace{-4mm}
\end{table}

\noindent\textbf{A5. Knob Fix.}
The \textit{Knob Fix} action performs targeted correction of invalid or conflicting configurations identified during validation. 
Given the current configuration $\mathcal{C}_{t-1}^{tuned}$, system and network Knowledge $\mathcal{K}^{sys}$, knob-level knowledge $\mathcal{K}^{knob}$, 
and the detected issues $\mathcal{E}_{t-1}^{val}$, this action generates a corrected configuration.
\begin{equation}
\mathcal{C}_t^{fix} = \mathcal{A}_{fix}(\mathcal{C}_{t-1}^{tuned}, \mathcal{K}^{sys}, \mathcal{K}^{knob}, \mathcal{E}_{t-1}^{val})
\end{equation}
where $\mathcal{C}_t^{fix}$ denotes the updated configuration. For the identified issues, the action adjusts the corresponding knobs based on configuration constraints, inter-knob dependencies, and system context. 
The correction focuses on resolving conflicts and restoring feasibility. 
After applying the fixes, the updated configuration is rechecked to ensure logical consistency and deployability. 
This action ensures that the configuration remains valid and executable, providing a reliable basis for subsequent tuning steps.
\vspace{1mm}

\noindent\textbf{A6. Performance Evaluation.}
The \textit{Performance Evaluation} action executes the current configuration $\mathcal{C}_{t-1}$ on the blockchain system to obtain its performance.
Given the system $\mathcal{B}$ and workload $S$, this action runs benchmark workloads and computes the resulting throughput. The resulting performance $T_t$ captures the effect of the applied tuning decisions, and is used to evaluate the current configuration, guide subsequent action selection, and support pruning of suboptimal configurations during the search.
\begin{equation}
T_t = \mathrm{Eval}(\mathcal{B}, \mathcal{C}_{t-1}, S)
\end{equation}

\noindent\textbf{A7. Feedback-driven Refinement.}
The \textit{Feedback-driven Refinement} action performs iterative refinement based on runtime feedback from previous evaluations. 
Unlike one-shot configuration generation, this action improves configurations through repeated observation and adjustment. 
Given the latest evaluation results, including performance metrics $T_t$ and execution errors $\mathcal{E}_t^{run}$, the current configuration $\mathcal{C}_{t}^{tuned}$, system and network knowledge $\mathcal{K}^{sys}$, and knob-level knowledge $\mathcal{K}^{knob}$, 
this action analyzes potential causes of performance degradation or failures and identifies bottleneck-related knobs for further refinement.
\begin{equation}
\mathcal{C}_t^{fb} = \mathcal{A}_{feedback}(\mathcal{C}_{t}^{tuned}, T_t, \mathcal{E}_t^{run}, \mathcal{K}^{knob}, \mathcal{K}^{sys})
\end{equation}
where $\mathcal{C}_t^{fb}$ denotes the updated configuration after incorporating feedback from the latest evaluation.
The action analyzes feedback to identify performance bottlenecks, selectively refines performance-limiting knobs while preserving effective settings, and re-evaluates the updated configuration to guide subsequent refinement.
This process continues until performance improvement becomes marginal or a maximum number of refinement iterations is reached. This design enables targeted refinement of bottleneck knobs and progressively improves configuration quality.
\vspace{1mm}

\noindent\textbf{A8. Terminal.}
The \textit{Terminal} action indicates the termination of the tuning process. It is triggered when predefined stopping conditions are met, such as the completion of feedback iterations or reaching the maximum search depth. Once activated, no further tuning actions are performed.
\vspace{1mm}

% \noindent\textbf{Action Ordering and Constraints.}
To constrain the search process, BCTuner defines explicit transition rules over the action space, restricting which actions can follow a given action. These rules enforce a valid ordering among actions and ensure that each action is applied at an appropriate stage of the tuning process (e.g., validation and fixing follow configuration adjustments, while feedback tuning is triggered only after evaluation). Table~\ref{tab:action_transition} summarizes the valid action transitions. These constraints guide the search into a structured progression, improving both efficiency and stability.

\subsection{MCTS-based Configuration Generation}
% \noindent\textbf{Configuration Generation with MCTS Rollout.}
% To generate high-quality configurations in a large and interdependent knob space, BCTuner employs MCTS to guide configuration generation. Instead of predicting a complete configuration in a single step, MCTS explores the search space incrementally through a sequence of structured actions, balancing exploration of new decisions and exploitation of promising configurations.

% In BCTuner, the search operates over the action space defined in Section~\ref{sec:action_design}, where each action transforms the current configuration into a new state. These states and transitions form a search tree, with nodes representing intermediate configurations and edges corresponding to valid tuning actions. Based on this formulation, MCTS iteratively explores and evaluates action sequences to identify promising tuning trajectories. The search follows the standard MCTS procedure, consisting of four stages: selection, expansion, simulation, and backpropagation as summarized in Algorithm~\ref{alg:mcts_generation}.

\noindent\textbf{Configuration Generation with MCTS Rollout.}
To generate configurations in a large and interdependent knob space, BCTuner employs MCTS to guide the search. Instead of predicting a complete configuration in one step, MCTS incrementally explores the space through structured actions, balancing exploration of new decisions and exploitation of promising configurations. In BCTuner, the search operates over the action space defined in Section~\ref{sec:action_design}. Each action transforms the current configuration into a new state, forming a search tree where nodes represent intermediate configurations and edges correspond to valid tuning actions. MCTS then explores and evaluates action sequences to identify promising tuning trajectories through four stages: selection, expansion, simulation, and backpropagation, as summarized in Algorithm~\ref{alg:mcts_generation}.

\begin{algorithm}[t]
\caption{MCTS-based Configuration Generation in BCTuner}
\label{alg:mcts_generation}

\begingroup
% \small
\setstretch{1.15}

\begin{algorithmic}[1]
\State Initialize the root state $s_0$ and search tree $\mathcal{T}$
\State $C^* \gets C(s_0)$, $T^* \gets benchmark(BC, C(s_0), S)$

\For{$i = 1$ to $maxRollout$}
    \State $v \gets s_0$, $\tau \gets \emptyset$

    \Statex \textbf{Selection}
    \While{$v$ is expanded and not terminal}
        \State $v \gets \arg\max_{a \in A(v)} 
        \left(
        \frac{Q(v,a)}{N(v,a)} + c \sqrt{\frac{\ln N(v)}{N(v,a)}}
        \right)$
        % \State Select $a^*$ by UCT and move to the corresponding child
    \EndWhile

    \Statex \textbf{Expansion}
    \If{$v$ is not terminal}
        \State Expand $v$ via valid actions and obtain a successor state
        \State $\tau \gets \tau \cup (v,a)$
    \EndIf

    \Statex \textbf{Simulation}
    \While{$v$ is not terminal}
        \State Execute the next valid action and update $v$
        \State $\tau \gets \tau \cup (v,a)$
    \EndWhile

    \Statex \textbf{Backpropagation}
    \State Compute reward $r$ for the rollout result
    \State Update $C^*$ and $T^*$ if a better configuration is found

    \ForAll{$(v_j,a_j) \in \tau$}
        \State Update $Q(v_j,a_j)$ and $N(v_j,a_j)$
    \EndFor
\EndFor

\State \Return $C^*$
\end{algorithmic}

\endgroup
\end{algorithm}
% \vspace{-4mm}

\vspace{1mm}
\noindent\textbf{Selection.} During selection, the search traverses the tree from the root node to either an unexpanded node or a terminal node, identifying promising nodes for further expansion. We adopt the Upper Confidence Bound applied to Trees (UCT) strategy~\cite{kocsis2006bandit} to balance exploration and exploitation during node selection. The UCT score of selecting action is defined as:
\begin{equation}
\mathrm{UCT}(v, a) = \frac{Q(v,a)}{N(v,a)} + c \cdot \sqrt{\frac{\ln N(v)}{N(v,a)}},
\end{equation}
where $N(v,a)$ denotes the number of times action $a$ has been selected from node $v$, and $N(v)$ is the total number of visits to node $v$. $Q(v,a)$ represents the accumulated reward of taking action $a$ from node $v$, which is updated during backpropagation. At each step, the action with the highest UCT score is selected to traverse the tree. If there exist unvisited child nodes (i.e., $N(v,a)=0$), they are prioritized to encourage exploration.

\vspace{1mm}
\noindent\textbf{Expansion.} During expansion, the selected node is expanded through valid actions, where the valid action set is determined by the current action type and the transition rules in Table~\ref{tab:action_transition}. For actions involving LLM reasoning, BCTuner invokes the LLM to generate executable action instances conditioned on the current tuning context, and samples each action multiple times to produce diverse child nodes.

For non-LLM actions (e.g., Performance Evaluation), the transition produces a deterministic successor. Since these actions interact with the real system and provide contextual information for later decisions, BCTuner replicates the resulting state to match the number of child nodes generated by LLM-based actions. In this way, each action contributes a consistent number of child nodes, allowing evaluation results to be effectively incorporated into subsequent search decisions and refinement steps.

\vspace{1mm}
\noindent\textbf{Simulation.} The simulation phase proceeds by repeatedly extending nodes through successive action selection and expansion steps, until reaching a termination node. All intermediate nodes generated during this process are retained in the search tree, enabling future iterations to build upon previously explored trajectories.

\vspace{1mm}
\noindent\textbf{Backpropagation.}
During backpropagation, the reward obtained from simulation is propagated backward along the visited trajectory to update the statistics of the corresponding nodes. Specifically, for each node-action pair $(v,a)$ along the trajectory $v_0 \oplus \cdots \oplus v_t$, the accumulated reward and visit count are updated as $Q(v,a) = Q(v,a) + r$ and $N(v) = N(v) + 1$. These updated statistics are then used to guide subsequent selection via the tree policy.

A key component of BCTuner is a reward mechanism tailored to different types of tuning actions, enabling effective guidance of the search process. For validation-related actions, a sparse reward is adopted to enforce feasibility constraints: an invalid configuration receives a negative reward $r=-1$, while a valid configuration is assigned $r=1$, allowing the search to quickly eliminate infeasible branches. For performance-related actions, including evaluation and feedback tuning, the reward is defined based on the relative performance improvement with respect to a baseline configuration:
\begin{equation}
r_T =
\left\{
\begin{aligned}
\phantom{-}\,e^{\frac{T - T_{\text{default}}}{T_{\text{default}}}}, \quad & T \ge T_{\text{default}}, \\[6pt]
-\,e^{\frac{T - T_{\text{default}}}{T_{\text{default}}}}, \quad & T < T_{\text{default}},
\end{aligned}
\right.
\end{equation}
where $T$ and $T_{default}$ denote the throughput of the current configuration and default configuration, respectively. Evaluation actions directly use this reward, while feedback-driven tuning follows the same formulation but further amplifies it as $r_{\text{feedback}} = 1.5 \cdot r_T$.This reward design integrates feasibility constraints with continuous performance signals, enabling the search to both prune invalid configurations and prioritize trajectories that lead to consistent performance improvement.

\vspace{1mm}
\noindent\textbf{Stopping Criteria.}
The overall search process terminates when one of the following conditions is satisfied: (1) a configuration achieves the predefined target throughput, or (2) the rollout budget is fully exhausted. Upon termination of the search, BCTuner returns the configuration with the highest throughput observed during the entire exploration process, among all candidates. 

\subsection{Pruning Strategy}

Although MCTS provides a framework for exploring the action space, expanding all valid successors can produce many low-value branches in blockchain knob tuning, where each evaluation requires deployment and benchmarking. To improve efficiency, BCTuner applies a rule-based pruning strategy during expansion. Instead of treating all actions equally, the pruning mechanism narrows the candidate set based on the search stage, action outcomes, and runtime feedback, removing actions that are premature, redundant, or unlikely to improve the search outcome. In addition, BCTuner applies a separate pruning rule in the feedback tuning stage to terminate refinement when further iterations become ineffective.

At a high level, the pruning strategy follows three principles. First, it enforces \emph{stage-aware progression}, introducing expensive or restrictive actions only after sufficient exploration. The search is initially restricted to cluster-level tuning at the optimization-direction stage, and the full action space is opened after cluster exploration. Validation is postponed until enough tuning actions have accumulated, while evaluation is enabled after sufficient adjustment and suppressed if it appears too frequently in nearby ancestors, preventing premature validation or evaluation.

Second, BCTuner applies \emph{outcome-aware pruning} after validation and evaluation. During validation, branches with unreasonable configurations are terminated when similar results appear in recent ancestors; otherwise, only the fix action is retained before further exploration. When validation succeeds, successors are restricted to adjustment or evaluation. During evaluation, branches are terminated if throughput drops below 90\% of the initial performance.

Third, BCTuner adopts \emph{performance-guided continuation rules} for evaluation and feedback. After evaluation, a branch enters feedback tuning only if its throughput reaches at least 80\% of the interval between the baseline and the best observed performance; otherwise, the search returns to the optimization-direction stage. Within feedback tuning, refinement is terminated if performance degrades by more than 10\%. Overall, pruning combines search-stage constraints, action outcomes, and runtime feedback to guide the search toward promising trajectories and reduce unnecessary evaluations.

%% file: 4-exp.tex
\section{EXPERIMENTAL EVALUATION}

\subsection{Experimental Setup}
% \textbf{Blockchain Platforms:} In this study, we select two representative blockchain systems for experimental evaluation: Hyperledger Fabric and ChainMaker. For Hyperledger Fabric, we adopt the latest LTS version v2.5.12, with the Raft consensus protocol enabled and the endorsement policy configured as "OR". All Fabric-related components are deployed and executed using Docker containers. For ChainMaker, we use version v2.3.7, where blockchain nodes are directly deployed and run on physical servers, while smart contracts are executed within a Docker-based container environment. All experimental configurations strictly follow the respective production deployment guidelines of each platform.
\textbf{Blockchain Platforms:} We evaluate two representative blockchain systems, Hyperledger Fabric (hereafter referred to as Fabric) and ChainMaker. For Fabric, we use the latest LTS version v2.5.12 with the Raft consensus protocol and an ``OR'' endorsement policy, deploying all components via Docker containers. For ChainMaker (v2.3.7), smart contracts are executed within a Docker-based containerized runtime environment.

% \noindent\textbf{Hardware and Environment:} All experiments are conducted in a cloud environment. The blockchain system consists of six cloud server instances, each provisioned with 16 vCPU cores based on AMD EPYC 9754 processors, 32 GB of memory, and 500 GB of SSD storage, running Ubuntu 22.04 LTS. Machines used for transaction workload generation and performance metric collection are also deployed in the cloud, each with 20 vCPU cores based on Intel Xeon Platinum 8576C processors, 96 GB of memory, and 1 TB of SSD storage. All servers and test clients are interconnected via a 5 Gbps high-speed internal network to minimize the impact of network bandwidth on experimental results. All tuning and control algorithms are implemented in Python 3.9.

\vspace{1mm}
\noindent\textbf{Hardware and Environment:} 
All experiments are conducted on cloud servers running Ubuntu 22.04 LTS. The blockchain system runs on six instances (16 vCPU, AMD EPYC 9754, 32 GB RAM, 500 GB SSD). Benchmark execution is performed on a separate client machine (Intel Xeon Platinum 8576C, 20 vCPU, 96 GB RAM, 1 TB SSD). All servers and clients are connected via a 5 Gbps internal network. BCTuner is implemented in Python 3.9.

% \noindent\textbf{Workload:}On Hyperledger Fabric, we use the Simple and SmallBank smart contracts as workloads, which have been widely adopted in prior studies and are considered representative of general-purpose computation and financial transaction processing in blockchain systems~\cite{10.14778/3579075.3579076, sharma2019blurring, gorenflo2020fastfabric}. Experiments are conducted using the official benchmarking tool Hyperledger Caliper, which submits a total of 50,000 transactions to the system under the fixed-rate mode. On ChainMaker, we evaluate the system using the Fact contract, a typical notarization-oriented smart contract for assessing read and write performance, and conduct experiments with the official tool chainmaker-bench, which also submits 50,000 transactions in automated testing mode. The reported TPS (transactions per second) measures the number of transactions successfully committed to the blockchain.
\vspace{1mm}
\noindent\textbf{Workload:} 
On Fabric, we use the \textit{Simple} and \textit{SmallBank} smart contracts, which are widely adopted and representative of general-purpose and financial workloads~\cite{10.14778/3579075.3579076, sharma2019blurring, gorenflo2020fastfabric}. Experiments are conducted using Hyperledger Caliper~\cite{caliper2026} with 50,000 transactions under fixed-rate mode. On ChainMaker, we evaluate the system using chainmaker-bench with 50,000 transactions in automated testing mode. TPS (transactions per second) measures the number of successfully committed transactions.

\vspace{1mm}
\noindent\textbf{Baseline:} We consider the following baseline methods:
\begin{itemize}[leftmargin=1.2em, itemsep=2pt, topsep=2pt]
  \item \textbf{SMAC}~\cite{hutter2011sequential} is a Bayesian optimization method that uses random forests to handle high-dimensional heterogeneous knob spaces.
  \item \textbf{GP}~\cite{duan2009tuning, van2017automatic} is a Bayesian optimization method that uses Gaussian processes to model configuration–performance relationships.
  \item \textbf{GPTuner}~\cite{lao2025gptuner} is an LLM-assisted tuning method that reduces the search space via knowledge extraction and applies coarse-to-fine Bayesian optimization.
  \item \textbf{Athena}~\cite{10.14778/3579075.3579076} is a reinforcement learning–based tuning system for Fabric that uses PB-MADDPG, a multi-agent deep reinforcement learning algorithm, to optimize system configurations.
\end{itemize}

% \noindent\textbf{Tuning Settings:}In all experiments, BCTuner jointly tunes all 120 configurable knobs in Fabric, with knob ranges automatically determined by BCTuner based on the official configuration manuals and knob semantic knowledge. For the other baseline methods, directly tuning the full knob set frequently generates configurations that violate inter-knob dependencies or constraints during the search process, resulting in a large number of invalid configurations and preventing the blockchain system from starting correctly. Therefore, based on expert knowledge, we select 50 knobs with the most significant impact on system performance for tuning. For these knobs, when value ranges are explicitly specified in the official documentation, we adopt them directly; otherwise, the lower bound is set to one-tenth of the default value and the upper bound to ten times the default value.
\vspace{1mm}
\noindent\textbf{Tuning Settings:} 
In all experiments, BCTuner jointly tunes all 120 Fabric knobs, with ranges derived from official manuals and knob semantics. Directly applying baseline methods to the full Fabric knob set is often ineffective, because inter-knob dependencies frequently lead to invalid configurations that prevent the blockchain network from starting properly. These invalid trials provide no valid performance feedback, limiting the ability of data-driven methods to update their surrogate models or search policies. Therefore, we select 50 performance-critical knobs for the baselines based on expert experience. For ChainMaker, all methods tune the complete set of 38 system-provided knobs. Value ranges are taken from official documentation when available; otherwise, they are set to [0.1$\times$, 10$\times$] of the default values. Each experiment is repeated three times. We report the best TPS from the median run. In the plots, the solid lines denote the median TPS, and the shaded regions show the observed TPS fluctuation across the three runs.
% In all experiments, BCTuner tunes all 120 Fabric knobs, with ranges derived from official manuals and knob semantics. Since applying baselines to the full knob set often yields invalid configurations due to inter-knob dependencies and prevents system startup, we select 50 performance-critical knobs for baselines based on expert experience. For ChainMaker, all methods tune the complete set of 38 system-provided knobs. Value ranges follow official documentation when available, and otherwise use [0.1$\times$, 10$\times$] of the default values. Each experiment is repeated three times; we report the best TPS from the median run, while solid lines in the plots denote median TPS and shaded regions show observed TPS fluctuation across runs.

\subsection{Tuning Effectiveness and Efficiency}

% \begin{figure}[t]
% \centering
% \includegraphics[width=\linewidth]{figures/smallbank.png}
% \caption{Tuning performance comparison under the \textit{SmallBank} workload across different network architectures. }
% \Description{Line plots showing throughput versus iteration for BCTuner, Athena, GPTuner, SMAC, GP, and Default under the SmallBank workload. BCTuner converges faster and achieves higher throughput across all settings.}
% \label{fig:smallbank}
% \end{figure}
% \begin{figure}[t]
% \centering
% \includegraphics[width=\linewidth]{figures/simple.png}
% \vspace{-3mm}
% \caption{Tuning performance comparison under the \textit{Simple} workload across different network architectures.}
% \Description{Line plots showing throughput versus iteration for BCTuner, Athena, GPTuner, SMAC, GP, and Default under the SmallBank workload. BCTuner converges faster and achieves higher throughput across all settings.}
% \vspace{-4mm}
% \label{fig:simple}
% \end{figure}

\begin{figure}[t]
\centering
\includegraphics[width=\linewidth]{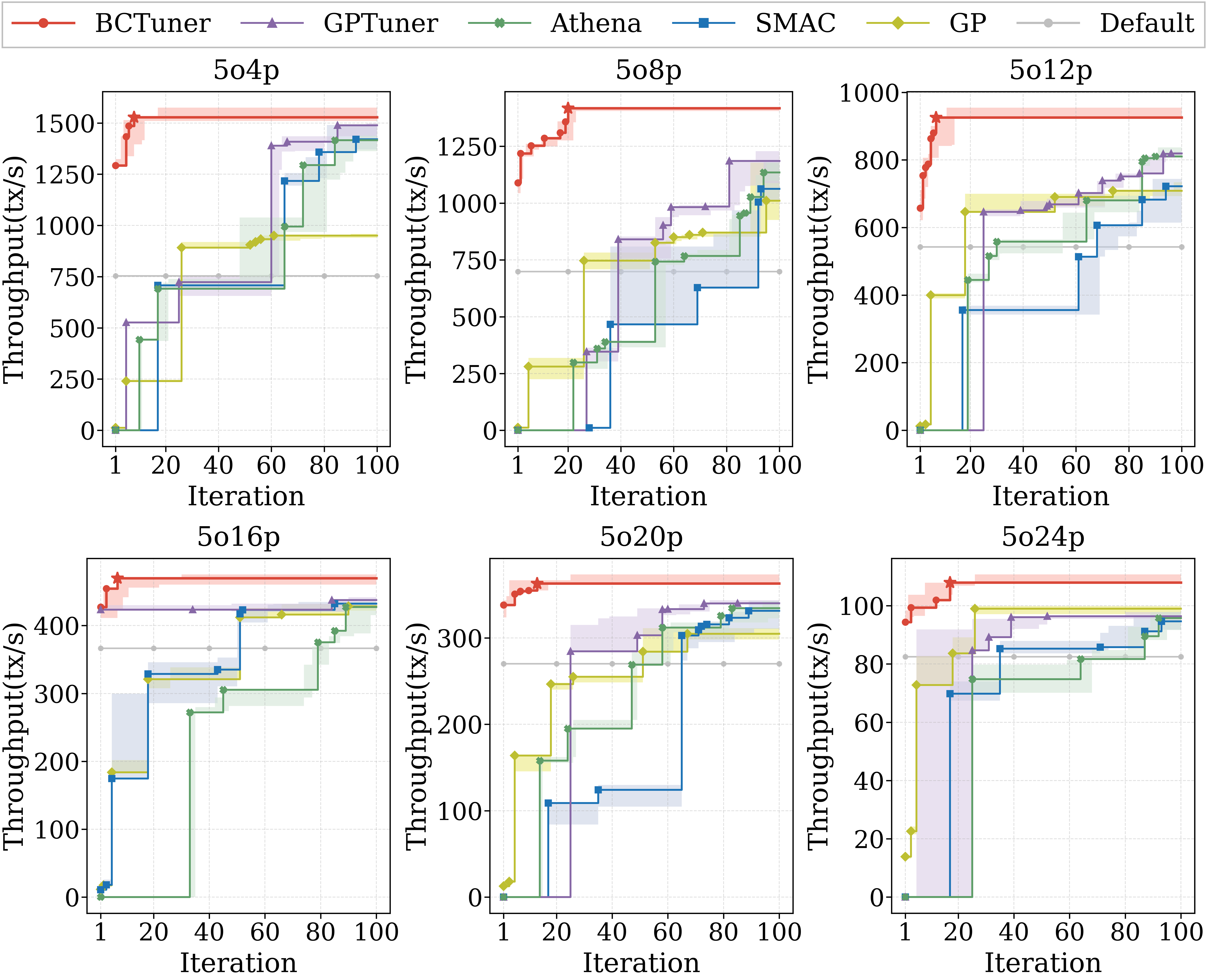}
% \vspace{-4mm}
\caption{Tuning performance comparison under the \textit{SmallBank} workload across different network architectures. }
\Description{Line plots showing throughput versus iteration for BCTuner, Athena, GPTuner, SMAC, GP, and Default under the SmallBank workload. BCTuner converges faster and achieves higher throughput across all settings.}
\vspace{-4mm}
\label{fig:smallbank}
\end{figure}

\begin{figure}[t]
\centering
\includegraphics[width=\linewidth]{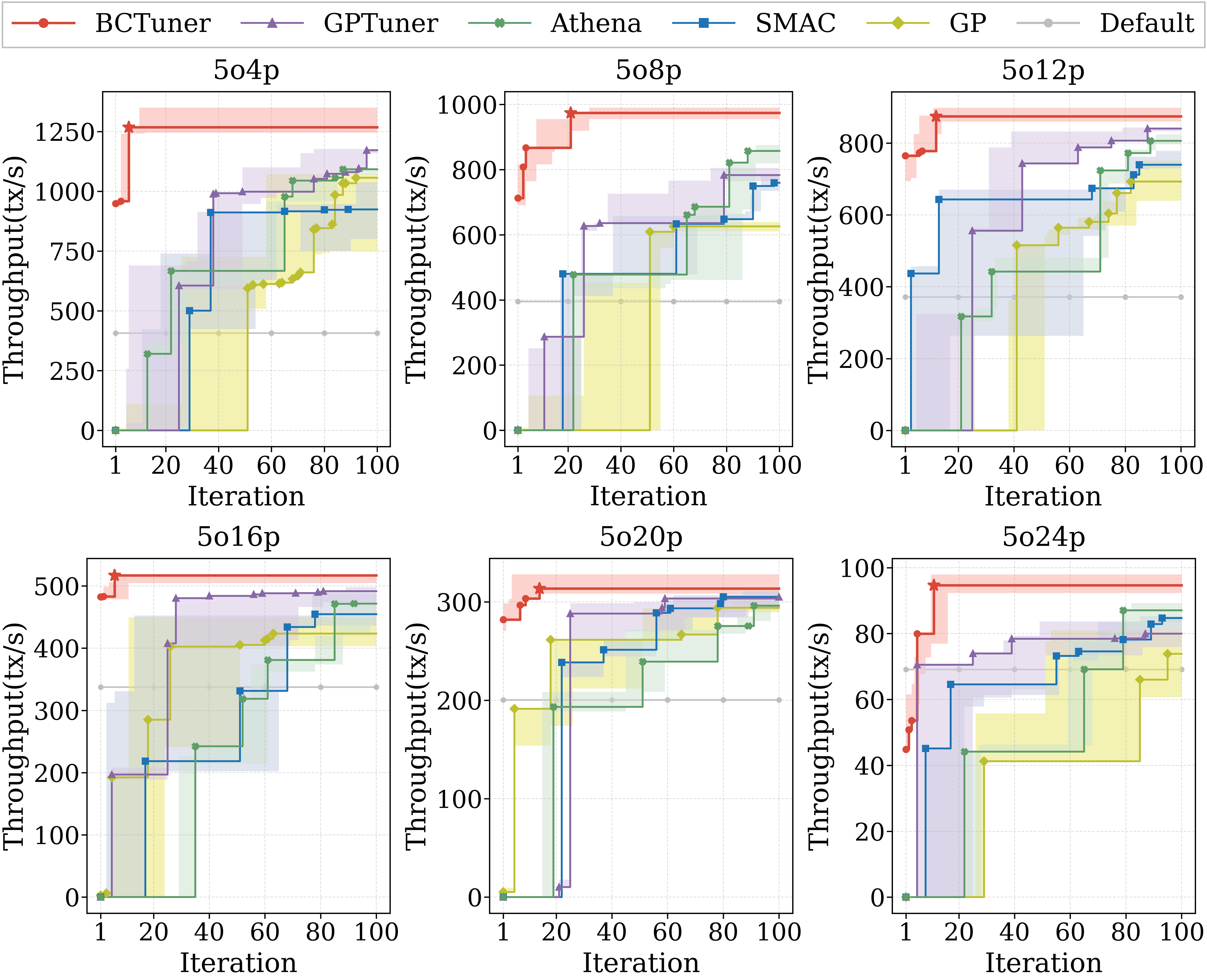}
% \vspace{-4mm}
\caption{Tuning performance comparison under the \textit{Simple} workload across different network architectures.}
\Description{Line plots showing throughput versus iteration for BCTuner, Athena, GPTuner, SMAC, GP, and Default under the SmallBank workload. BCTuner converges faster and achieves higher throughput across all settings.}
\vspace{-6mm}
\label{fig:simple}
\end{figure}

This subsection evaluates the proposed method \textbf{BCTuner} against \textbf{Default}, \textbf{GPTuner}, \textbf{Athena}, \textbf{SMAC}, and \textbf{GP} under different workloads and Fabric network architectures to assess its effectiveness and efficiency in knob tuning. We consider the \textit{SmallBank} and \textit{Simple} smart contracts as representative workloads and conduct experiments on Fabric networks with 4, 8, 12, 16, 20, and 24 Peer nodes. In all experiments, the number of Orderer nodes is fixed at five. Peer nodes are deployed with two Peers per organization across six servers, while Orderer nodes are placed on five of these servers. The results are shown in Fig.~\ref{fig:smallbank} and Fig.~\ref{fig:simple}.

% We compare \textbf{BCTuner} with \textbf{Default}, \textbf{GPTuner}, \textbf{Athena}, \textbf{SMAC}, and \textbf{GP} under different workloads and Fabric network architectures. Experiments are conducted using the \textit{SmallBank} and \textit{Simple} smart contracts on networks with 4, 8, 12, 16, 20, and 24 Peer nodes, with the number of Orderers fixed at five. Peer nodes are deployed with two nodes per organization across six servers, while Orderer nodes are distributed across five servers. The results are presented in Fig.~\ref{fig:smallbank} and Fig.~\ref{fig:simple}.

Overall, the results show that \textbf{BCTuner} consistently achieves the best performance across all experimental settings. Across different Fabric network architectures ranging from 4 to 24 Peer nodes, BCTuner achieves TPS improvements of (102.77\%, 102.84\%, 70.57\%, 28.2\%, 34.39\%, 30.87\%) in the \textit{SmallBank} workload over default settings, with an average gain of 61.61\%, and reaches peak performance within (8, 20, 7, 7, 13, 17) iterations. For the \textit{Simple} workload, BCTuner achieves TPS improvements of (211.38\%, 146.62\%, 135.4\%, 53.22\%, 56.45\%, 36.95\%), with an average gain of 106.67\%, and converges to the optimal configuration within (6, 21, 12, 6, 14, 11) iterations. These results suggest that BCTuner effectively uncovers performance potential left untapped by default configurations, delivering substantial throughput improvements across network scales while requiring relatively few iterations.

% Compared with \textbf{GPTuner}, BCTuner demonstrates clear advantages in both TPS performance and convergence efficiency. As shown in Fig.~4 and Fig.~5, across experiments with 4--24 Peer nodes, BCTuner achieves an average maximum TPS improvement of 14.88\% over GPTuner in the \textit{SmallBank} workload (5.26\%, 33.21\%, 19.56\%, 8.76\%, 8.45\%, 14.02\%), while reaching peak performance $7.33\times$ faster on average (9.63, 3.05, 12.71, 11, 5.5, 2.05). In the \textit{Simple} workload, BCTuner achieves an average TPS improvement of 19.01\% (23.52\%, 48.22\%, 9.17\%, 7.62\%, 4.33\%, 21.22\%) and converges $8.27\times$ faster on average (15, 2.76, 6.33, 12.5, 6.14, 6.91). This performance gap can be attributed to the fundamental differences in how the two methods utilize knowledge. GPTuner relies on historical knowledge to reduce the search space, and its optimization process remains constrained by the Bayesian optimization framework. However, in blockchain systems, reusable prior knowledge is relatively limited, and the complex interdependencies among knobs make it difficult for GPTuner’s single-knob-oriented knowledge utilization to capture global dependencies. In contrast, BCTuner more effectively leverages system-level knowledge extracted from manual and integrates it with a clustered search strategy to guide exploration, thereby achieving more efficient and effective tuning.

\vspace{1mm}
Compared with \textbf{GPTuner}, BCTuner achieves higher TPS and faster convergence across all settings. As shown in Fig.~\ref{fig:smallbank} and Fig.~\ref{fig:simple}, under 4--24 Peer nodes, BCTuner improves TPS by 14.88\% on average on \textit{SmallBank} (5.26\%, 33.21\%, 19.56\%, 8.76\%, 8.45\%, 14.02\%) with $7.33\times$ faster convergence (9.63, 3.05, 12.71, 11, 5.5, 2.05). On \textit{Simple}, it achieves 19.01\% improvement (23.52\%, 48.22\%, 9.17\%, 7.62\%, 4.33\%, 21.22\%) with $8.27\times$ faster convergence (15, 2.76, 6.33, 12.5, 6.14, 6.91).
This gap reflects differences in knowledge utilization and search strategy. GPTuner reduces the search space using prior knowledge but performs optimization within a Bayesian framework. In blockchain systems, reusable prior knowledge is limited, and complex interactions among knobs make it difficult for such approaches to capture global dependencies. In contrast, BCTuner incorporates knob knowledge from manuals and combines it with MCTS process, reducing ineffective trials and improving convergence and final performance.

\vspace{1mm}
\textbf{Athena} adopts an online reinforcement learning approach and relies on extensive system interactions, requiring a large number of trials to learn effective tuning strategies. Across all network scales, BCTuner achieves higher peak TPS and faster convergence on both \textit{SmallBank} and \textit{Simple} workloads. 
On \textit{SmallBank}, BCTuner improves TPS by 18.94\% on average (14.91\%, 40.48\%, 21.20\%, 11.62\%, 10.60\%, 14.82\%) and achieves $7.90\times$ faster convergence (9.5, 3.7, 11.85, 11.71, 6.23, 4.41). On \textit{Simple}, it achieves 20.71\% average improvement (42.21\%, 29.5\%, 18.33\%, 13.51\%, 8.7\%, 11\%) with $8.19\times$ faster convergence (13.5, 3.19, 6.41, 14.3, 5.5, 6.18). 
This gap mainly reflects differences in exploration efficiency. Reinforcement learning methods typically require extensive trial-and-error interactions to learn effective tuning policies. In contrast, BCTuner uses knob semantics, system-level knowledge, and structured tuning actions to constrain the search process, thereby reducing ineffective trials and achieving faster convergence with higher performance.

% Compared with traditional Bayesian optimization methods \textbf{SMAC} and \textbf{GP}, BCTuner demonstrates significant advantages in both performance and convergence efficiency. Due to the lack of system-level knowledge guidance, SMAC and GP tend to generate invalid configurations that prevent the system from running correctly, even when expert knowledge is used to restrict the search space. As shown in Fig.~4 and Fig.~5, in the \textit{SmallBank} workload, BCTuner achieves an average maximum TPS improvement of 23.42\% over SMAC (14.29\%, 50.72\%, 37.46\%, 10.21\%, 11.69\%, 16.14\%) and 36.43\% over GP (76.6\%, 58.24\%, 39.9\%, 11.29\%, 21.56\%, 10.95\%), while improving convergence speed by $8.01\times$ (10.5, 3.65, 12.43, 11.14, 5.85, 4.47) and $6.08\times$ (6.625, 3.75, 9.57, 11.86, 4.15, 0.529), respectively. In the \textit{Simple} workload, BCTuner achieves an average maximum TPS improvement of 35.37\% over SMAC (84.43\%, 54.34\%, 36.31\%, 18.56\%, 4.18\%, 14.38\%) and 41.5\% over GP (51.86\%, 88.08\%, 48.81\%, 27.72\%, 9.77\%, 22.77\%), while improving convergence speed by $7.96\times$ (13.83, 3.66, 6.08, 12, 4.71, 7.45) and $7\times$ (14.33, 1.86, 5.8, 9.5, 4.57, 5.9), respectively. These results further demonstrate that incorporating system-level knowledge significantly improves the effectiveness and stability of the search process, enabling BCTuner to avoid invalid configurations while consistently achieving superior performance, and thus outperform traditional Bayesian optimization methods overall.
\vspace{1mm}
Compared with \textbf{SMAC} and \textbf{GP}, BCTuner achieves higher TPS and faster convergence. As shown in Fig.~\ref{fig:smallbank} and Fig.~\ref{fig:simple}, on \textit{SmallBank}, BCTuner improves TPS by 23.42\% over SMAC (14.29\%, 50.72\%, 37.46\%, 10.21\%, 11.69\%, 16.14\%) and 36.43\% over GP (76.6\%, 58.24\%, 39.9\%, 11.29\%, 21.56\%, 10.95\%), with $8.01\times$ (10.5, 3.65, 12.43, 11.14, 5.85, 4.47) and $6.08\times$ (6.625, 3.75, 9.57, 11.86, 4.15, 0.529) faster convergence, respectively. On \textit{Simple}, it achieves 35.37\% improvement over SMAC (84.43\%, 54.34\%, 36.31\%, 18.56\%, 4.18\%, 14.38\%) and 41.5\% over GP (51.86\%, 88.08\%, 48.81\%, 27.72\%, 9.77\%, 22.77\%), with $7.96\times$ (13.83, 3.66, 6.08, 12, 4.71, 7.45) and $7\times$ (14.33, 1.86, 5.8, 9.5, 4.57, 5.9) faster convergence, respectively. 
This gap is mainly due to differences in search guidance. Without knob and system knowledge, Bayesian optimization relies on trial-based exploration, leading to inefficient search and slower convergence. In contrast, BCTuner guides the search using multi-source knowledge and structured tuning actions, improving search efficiency and performance.

\vspace{1mm}
Further analysis across different network scales shows that \textbf{BCTuner} exhibits more pronounced performance advantages over other methods in configurations with 4--12 Peer nodes. In contrast, under configurations with 16--24 Peer nodes, where some servers host both Orderer and multiple Peer nodes, system resources gradually become a bottleneck, limiting the overall room for performance improvement. Even under these conditions, BCTuner consistently outperforms all baselines, demonstrating strong robustness. Moreover, benefiting from its pruning mechanism, BCTuner can effectively filter candidate configurations before deployment, thereby reducing unnecessary evaluation overhead; this advantage is particularly evident under low-TPS conditions. Overall, by effectively leveraging domain knowledge and guiding the search process, BCTuner significantly outperforms existing methods in both tuning effectiveness and efficiency.

% \vspace{-3mm}
\subsection{Ablation Study}
% To systematically evaluate the contribution of each key component in the proposed framework, we conduct a set of ablation studies during the tuning process. In addition, we further analyze the impact of different LLMs on tuning performance. Unless otherwise specified, all ablation experiments are conducted on a 5o8p network configuration using the SmallBank workload. We use the following metrics in our evaluation: $\Delta T$ (performance improvement over the default configuration), $T^*$ (the best achieved TPS), $N^*$ (the number of evaluations required to reach $T^*$), $N_{\text{neg}}$ (the number of configurations with performance lower than the default), and $N_{\text{err}}$ (the number of invalid configurations).

% We conduct ablation studies to evaluate the contribution of key components in the proposed framework, and further analyze the impact of different LLMs on tuning performance. Unless otherwise specified, experiments are performed on a 5o8p network using the \textit{SmallBank} workload. We report the following metrics:$T^*$ (best achieved TPS), $\Delta T$ (performance improvement over the default configuration), $N^*$ (number of evaluations required to reach $T^*$), $N_{\text{neg}}$ (number of configurations with performance lower than the default), and $N_{\text{err}}$ (number of invalid configurations). 

The purpose of this ablation study is to analyze the contribution of key components in BCTuner, including multi-Source blockchain knowledge, tuning actions, and pruning strategy. All experiments are performed on a 5o8p network using the \textit{SmallBank} workload. We report the following metrics: $T^*$ (best achieved TPS), $\Delta T$ (relative improvement over the default configuration, in \%), $N^*$ (number of evaluations required to reach $T^*$), $N_{\text{neg}}$ (number of configurations with performance lower than the default), and $N_{\text{err}}$ (number of invalid configurations).

\begin{table}[t]
\centering
\caption{Ablation study of BCTuner under the 5o8p network with the SmallBank workload.}
% \vspace{-3mm}
\label{tab:ablation}
\renewcommand{\arraystretch}{1.15} % 行距变大
\setlength{\tabcolsep}{3.5pt}
\begin{tabular}{lccccc}
\toprule
% Method & $T^*$ & $\Delta T$ (\%) & $N^*$ & $N_{\text{neg}}$ & $N_{\text{err}}$ \\
% \textbf{Method} & \textbf{$T^*$} & \textbf{$\Delta T$ (\%)} & \textbf{$N^*$} & \textbf{$N_{\text{neg}}$} & \textbf{$N_{\text{err}}$} \\
\textbf{Method} & \boldmath$T^*$ & \boldmath$\Delta T$ (\%) & \boldmath$N^*$ & \boldmath$N_{\text{neg}}$ & \boldmath$N_{\text{err}}$ \\
\midrule
\textbf{BCTuner} & \textbf{1417.80} & \textbf{102.85} & \textbf{20} & \textbf{0/24} & \textbf{0/24} \\
% \midrule
% ===== Background Knowledge =====
- w/o Knob Knowl & 1174.25 & 68.00 & 9 & 6/17 & 8/17 \\
- w/o Hardware Knowl & 1375.55 & 96.80 & 29 & 8/31 & 2/31 \\
- w/o Network Knowl & 1281.71 & 83.38 & 11 & 0/13 & 0/13 \\
% \addlinespace[4pt]

% ===== Action Modules =====
- w/o Global Tuning Plan & 1360.04 & 94.58 & 21 & 0/26 & 2/26 \\
- w/o Cluster-wise Tuning & 1337.18 & 91.31 & 15 & 0/17 & 4/17 \\
- w/o Single-knob Tuning & 1356.88 & 94.13 & 34 & 0/34 & 0/34 \\
- w/o Validation \& Fix & 1360.04 & 94.58 & 21 & 0/26 & 2/26 \\
- w/o  Feedback-driven & 1227.73 & 75.65 & 23 & 3/30 & 1/30 \\
% \addlinespace[4pt]

% ===== Pruning =====
- w/o Pruning Rules & 1386.24 & 98.33 & 39 & 0/43 & 6/43 \\
\bottomrule
\end{tabular}
\vspace{-3mm}
\end{table}

\vspace{1mm}
\noindent\textbf{Impact of Multi-Source Blockchain Knowledge.}
We analyze the effect of removing different knowledge, including knob, hardware and network knowledge. As shown in Table~\ref{tab:ablation}, eliminating any of these knowledge sources leads to performance degradation.

% Removing knob knowledge results in the largest degradation, with $\Delta T$ dropping by 34.85\%, and significantly increases invalid configurations, with nearly half of the configurations being invalid. It also leads to many configurations performing worse than the default, indicating that knob-level semantics are necessary to ensure configuration validity and effective search.

% Removing hardware knowledge reduces tuning stability, with nearly one third of configurations either invalid or performing worse than the default. This indicates that without accurate knowledge of hardware capacity, configurations may violate system constraints or fail to utilize available resources effectively.

% Removing network knowledge decreases $\Delta T$ by 19.47\%. Although no invalid configurations are observed, the search converges to suboptimal regions and fails to further improve performance, indicating that network-level knowledge is necessary to capture system characteristics and guide the search.

Removing \textit{knob knowledge} causes the largest degradation, with $\Delta T$ dropping by 34.85\%, and leads to a high proportion of invalid configurations, with nearly half being invalid. It also produces many configurations worse than the default, indicating that knob-level semantics are essential for ensuring configuration validity and effective search. Removing \textit{hardware knowledge} reduces tuning stability, with nearly one third of configurations either invalid or worse than the default, suggesting that without accurate capacity information, configurations may violate hardware constraints or underutilize resources. Removing \textit{network knowledge} decreases $\Delta T$ by 19.47\%; although no invalid configurations are observed, the search converges to suboptimal regions, indicating that network-level knowledge is required to capture blockchain system characteristics and guide the search.

Overall, different types of knowledge play complementary roles in BCTuner. 
Knob knowledge ensures validity, hardware knowledge improves stability, and network knowledge guides the search toward better configurations. These results highlight that effective tuning requires the combined use of these knowledge sources.

\vspace{1mm}
\noindent\textbf{Impact of Tuning Actions.}
We analyze the effect of removing different action modules in the MCTS process, including Global Tuning Plan, Cluster-wise Tuning, Single-knob Tuning, Validation and Fix, and Feedback-driven Refinement.

% Removing Global Tuning Plan has limited impact on final performance, suggesting that its role can be partially compensated by subsequent actions during the search process.

% Removing Cluster-wise Tuning leads to a clear performance drop, with $\Delta T$ dropping by 11.5\%., and results in a high proportion of invalid configurations (4/17). Without structured grouping, the search operates over the full knob space, increasing dimensionality and making it difficult to capture interactions among knobs.

% Removing Single-knob Tuning increases $N^*$ from 20 to 34, indicating substantially slower convergence. Without fine-grained adjustments, the search is unable to efficiently refine configurations.

% Removing Validation \& Fix introduces invalid configurations (2/26), indicating reduced robustness. Without validation, infeasible configurations may propagate during the search, degrading reliability.

% Removing Feedback-driven Refinement causes substantial performance degradation, with $\Delta T$ dropping by 27.2\%. Without feedback, the LLM cannot incorporate execution errors and performance outcomes, leading to less targeted updates, weakening the alignment between actions and observed outcomes.

% Overall, these actions play complementary roles in structuring and stabilizing the search process. Cluster-wise Tuning and Feedback-driven Refinement have the largest impact on performance, while Single-knob Tuning improves efficiency and Validation \& Fix ensures robustness.

Removing \textit{Global Tuning Plan} has limited impact on final performance, suggesting that its role can be partially compensated by subsequent actions during the search process. In contrast, removing \textit{Cluster-wise Tuning} leads to a clear performance drop, with $\Delta T$ decreasing by 11.5\%, and results in a high proportion of invalid configurations (4/17). Without structured grouping, the search operates over the full knob space, increasing dimensionality and making it difficult to capture interactions among knobs. Removing \textit{Single-knob Tuning} increases $N^*$ from 20 to 34, indicating substantially slower convergence, as the search cannot efficiently refine configurations without fine-grained adjustments. Removing \textit{Validation \& Fix} introduces invalid configurations (2/26), indicating reduced robustness, since infeasible configurations may propagate during the search and degrade reliability. Removing \textit{Feedback-driven Refinement} causes substantial performance degradation, with $\Delta T$ dropping by 27.2\%. Without feedback, the LLM cannot incorporate execution errors and performance outcomes, weakening the alignment between actions and observed outcomes.

Overall, these actions play complementary roles in structuring and stabilizing the search process. Cluster-wise Tuning and Feedback-driven Refinement have the largest impact on performance, while Single-knob Tuning improves efficiency and Validation \& Fix ensures robustness.

% \vspace{2mm}
% \noindent\textbf{Impact of Pruning Rules.}
% Removing the pruning rules is implemented by disabling the pruning strategy during the search process. 
% This leads to a substantial increase in the number of evaluations, with the search expanding from 24 to 43 steps, and requiring 39 evaluations to reach the best configuration. 
% Without pruning, the search is forced to explore a significantly larger portion of the configuration space, including many redundant and low-quality candidates, which markedly reduces search efficiency. 
% These results highlight that pruning is essential for constraining the search space and improving the efficiency of exploration in high-dimensional tuning scenarios.

\vspace{1mm}
\noindent\textbf{Impact of Pruning Rules.}
Disabling pruning during the search leads to a substantial increase in evaluations. 
The search expands from 24 to 43 steps and requires 39 evaluations to reach the best configuration. 
Without pruning, the search explores a larger portion of the configuration space, including redundant and low-quality candidates, which reduces efficiency. 
These results indicate that pruning is essential for constraining the search space and improving exploration efficiency in high-dimensional tuning.

\vspace{-2mm}
\subsection{Cross-System Adaptability}
\begin{figure}[t]
\centering
\includegraphics[width=\linewidth]{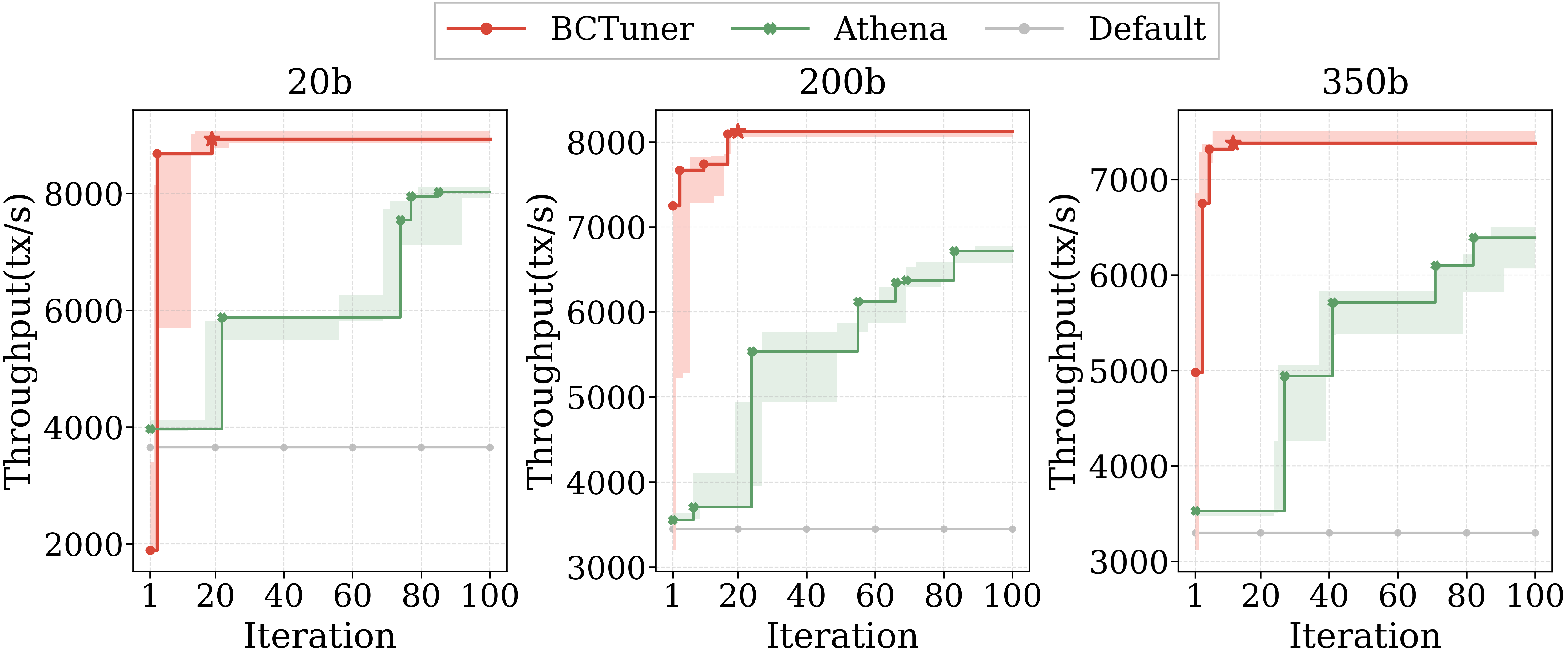}
\caption{Tuning performance comparison on ChainMaker under different workloads.}
\label{fig:chainmaker}
\vspace{-4mm}
\end{figure}

To evaluate the adaptability of BCTuner across blockchain systems, we conduct experiments on ChainMaker, which differs from Fabric in both architecture and execution model. Fabric follows an execute--order--validate paradigm with modularized stages, while ChainMaker adopts a more integrated transaction processing pipeline with distinct consensus and execution mechanisms, resulting in different performance bottlenecks and tuning behaviors.

BCTuner consistently outperforms the default configuration across all workloads. Under 20p, 200b, and 350b, it achieves TPS improvements of 144.58\%, 135.48\%, and 123.71\%, respectively, with an average gain of 134.59\%, and reaches the best configuration within 19, 20, and 12 evaluations. Compared with Athena, BCTuner achieves higher TPS in all settings, with improvements of 11.22\%, 20.94\%, and 15.48\%. It also converges $4.15\times$ faster on average, requiring fewer evaluations to reach high-quality configurations. As shown in Fig.~\ref{fig:chainmaker}, BCTuner attains higher performance with fewer iterations, while Athena converges more slowly and often remains in suboptimal regions.

These results demonstrate the adaptability of BCTuner across different blockchain systems. By leveraging knowledge derived from system documentation and incorporating runtime feedback, BCTuner captures system-specific characteristics and adjusts configurations accordingly. Combined with MCTS-based search, it achieves effective tuning without system-specific customization.

\begin{table*}[t!]
\centering
\caption{Time Breakdown of BCTuner under Different Workloads and Network Architectures}
\vspace{-2.8mm}
\label{tab:time_breakdown}
\renewcommand{\arraystretch}{1.15}
\resizebox{\textwidth}{!}{
\begin{tabular}{cccccccccccccc}
\toprule
% Workload & Network & All & SR & DP & EV & A1 & A2 & A3 & A4 & A5 & A6 & A7 & Overhead (\%) \\
\textbf{Workload} & \textbf{Network} & \textbf{All} & \textbf{SR} & \textbf{DP} & \textbf{EV} & \textbf{A1} & \textbf{A2} & \textbf{A3} & \textbf{A4} & \textbf{A5} & \textbf{A6} & \textbf{A7} & \textbf{Overhead (\%)} \\
\midrule
\textit{SmallBank} & 5o8p  & 4615.89 & 345.38 & 1478.42 & 2956.80 & 4.62 & 4.70 & 4.29 & 5.77 & 7.42 & 244.27 & 453.96 & \textbf{7.48} \\
\textit{SmallBank} & 5o20p & 11283.31 & 345.68 & 4102.36 & 6872.30 & 4.35 & 4.57 & 4.03 & 5.41 & 8.18 & 259.41 & 1558.06 & \textbf{3.06} \\
\textit{Simple}    & 5o8p  & 7307.65 & 368.88 & 2288.86 & 4750.87 & 2.91 & 5.34 & 3.72 & 5.46 & 5.70 & 229.73 & 1160.29 & \textbf{5.05} \\
\textit{Simple}    & 5o20p & 12435.49 & 337.25 & 4244.48 & 8036.89 & 2.59 & 4.75 & 3.76 & 5.83 & 6.04 & 604.50 & 2228.73 & \textbf{2.71} \\
\bottomrule
\end{tabular}
}
\end{table*}

\begin{table*}[t]
\centering
\caption{Comparison of Different LLM Backbones under the 5o8p Network with the \textit{SmallBank} Workload}
\vspace{-2.8mm}
\label{tab:llm_comparison}
\setlength{\tabcolsep}{6.8pt}
\renewcommand{\arraystretch}{1.15}
\begin{tabular}{lcccccccccc}
\toprule
\textbf{LLM Backbone} & \boldmath$T^*$ & \boldmath$\Delta T$ & \boldmath$N^*$ & \boldmath$N_{\text{neg}}$ & \boldmath$N_{\text{err}}$ & \textbf{Interact} & \textbf{Prompt Tokens} & \textbf{Completion Tokens} & \textbf{Cost (\$)} \\
\midrule
GPT-5-mini           & 1417.80 & 102.85 & 20 & 0/24 & 0/24 & 186 & 2952.87K & 50.04K & 0.84 \\
GPT-5                & 1435.68 & 105.41 & 15 & 0/18 & 2/18 & 205 & 3189.12K & 50.47K & 4.49 \\
GPT-5.2              & 1433.75 & 105.13 & 15 & 2/18 & 0/18 & 164 & 2596.24K & 31.06K & 4.98 \\
Gemini-3-Flash       & 1399.00 & 100.16 & 18 & 0/24 & 0/19 & 198 & 3037.29K & 22.46K & 1.59 \\
Gemini-3.1-Pro       & \textbf{1445.56} & \textbf{106.82} & 15 & 0/18 & 0/18 & 284 & 4441.36K & 37.79K & 9.34 \\
Gemini-2.5-Flash     & 1130.83 & 61.79  & 14 & 1/15 & 1/15 & 148 & 2309.42K & 29.71K & \textbf{0.77} \\
Gemini-2.5-Pro       & 1369.72 & 95.97  & 10 & 0/18 & 4/18 & 157 & 2429.89K & 19.94K & 3.24 \\
Deepseek-v3          & 1078.75 & 54.34  & 3  & 0/12 & 6/12 & 210 & 3228.82K & 26.01K & 0.95 \\
Deepseek-R1          & 1333.69 & 90.81  & 18 & 2/26 & 1/26 & 199 & 3256.69K & 33.08K & 0.96 \\
Qwen3-32b            & 939.93  & 34.48  & 4  & 0/8  & 3/8  & 167 & 2516.29K & 12.13K & - \\
\bottomrule
\end{tabular}
\vspace{-2.5mm}
\end{table*}

\vspace{-2mm}
\subsection{Execution Time Breakdown}
% To analyze the tuning time of BCTuner, we introduce a set of time-related metrics. Total tuning time (All) consists of search overhead (SR), deployment (DP), and evaluation (EV). DP denotes the time to deploy and initialize the blockchain system under a given configuration, EV denotes workload execution and metric collection, and SR denotes the computational overhead of MCTS, including node selection and backpropagation. Action execution (AE) is decomposed into A1--A7 (Direction, ClusterAdjustment, KnobFineTuning, KnobValidation, KnobFix, ShortEvaluation, and FeedbackEvaluation), each reported as the average execution time per step.
To analyze the tuning time of BCTuner, we introduce a set of time-related metrics. \textbf{Total tuning time (All)} consists of \textbf{search overhead (SR)}, \textbf{deployment (DP)}, and \textbf{evaluation (EV)}. \textbf{DP} denotes the time to deploy and initialize the blockchain system under a given configuration, \textbf{EV} denotes workload execution and metric collection, and \textbf{SR} denotes the computational overhead of MCTS, including node selection and backpropagation. \textbf{Action execution (AE)} is decomposed into A1--A7, corresponding to \textit{Global Tuning Plan}, \textit{Cluster-wise Tuning}, \textit{Single-knob Tuning}, \textit{Knob Validation}, \textit{Knob Fix}, \textit{Performance Evaluation}, and \textit{Feedback-driven Refinement}, each reported as the average execution time per step.

As shown in Table~\ref{tab:time_breakdown}, we report both the overall time breakdown and the average execution time of A1--A7. The results show that DP and EV dominate the total tuning time across workloads (\textit{SmallBank} and \textit{Simple}) and network settings (5o8p and 5o20p), while SR remains small, indicating that the main bottleneck lies in deployment and benchmarking rather than the search process. Among actions, A6 and A7 incur higher costs due to their reliance on deployment and workload evaluation.

BCTuner improves efficiency by reducing the number of deployment and evaluation rounds through guided search, enabling it to reach high-quality configurations with fewer evaluations and lower overall tuning time. This advantage is more pronounced in low-TPS settings, where each evaluation is more expensive. Overall, BCTuner reduces costly system operations, improving efficiency and scalability in practical settings.

\vspace{-2mm}
\subsection{Effect of Different LLM Backbones}

We further study how the choice of LLM backbone affects BCTuner by conducting experiments on the 5o8p Fabric network with the SmallBank workload. We instantiate BCTuner with ten models, including GPT-5-mini, GPT-5, GPT-5.2, Gemini-3-Flash, Gemini-3.1-Pro, Gemini-2.5-Flash, Gemini-2.5-Pro, Deepseek-v3, Deepseek-R1, and Qwen3-32b. As shown in Table 4, all evaluated backbones improve throughput over the default configuration, while exhibiting clear differences in final performance, convergence, invalid configurations, token usage, and monetary cost. Gemini-3.1-Pro achieves the highest throughput, while GPT-5 and GPT-5.2 obtain comparable performance with fewer evaluations. GPT-5-mini and Gemini-2.5-Flash achieve competitive tuning performance at relatively low monetary cost. Overall, these results show that BCTuner can work with different LLM backbones, and its effectiveness is not tied to a specific model.

\vspace{-2mm}
\subsection{Cost analysis}

In LLM-driven blockchain knob tuning, the overhead mainly comes from LLM interactions. BCTuner avoids costly data collection and pre-training by leveraging textual knowledge, and reaches $N^*$ within few iterations. We evaluate cost in terms of token consumption (prompt and completion), monetary expense, and interaction steps. As shown in Table~\ref{tab:llm_comparison}, total token usage ranges from 2.3M to 4.4M across models, corresponding to \$0.77--\$9.34. Efficiency varies across backbones: lightweight models (e.g., GPT-5-mini and Gemini-2.5-Flash) achieve comparable performance with lower token usage. Overall, BCTuner maintains low LLM-related cost while achieving strong tuning performance, indicating good cost-effectiveness for practical deployment.